\documentclass[twocolumn]{IEEEtran}
\usepackage[T1]{fontenc}
\usepackage[latin9]{inputenc}
\usepackage{verbatim}
\usepackage{float}
\usepackage{textcomp}
\usepackage{amsmath}
\usepackage{amssymb}
\usepackage{graphicx}
\usepackage[unicode=true,
 bookmarks=true,bookmarksnumbered=true,bookmarksopen=true,bookmarksopenlevel=1,
 breaklinks=false,pdfborder={0 0 0},pdfborderstyle={},backref=false,colorlinks=false]
 {hyperref}
\hypersetup{pdftitle={Your Title},
 pdfauthor={Your Name},
 pdfpagelayout=OneColumn, pdfnewwindow=true, pdfstartview=XYZ, plainpages=false}

\makeatletter

\newcommand{\lyxmathsym}[1]{\ifmmode\begingroup\def\b@ld{bold}
  \text{\ifx\math@version\b@ld\bfseries\fi#1}\endgroup\else#1\fi}

\providecommand{\tabularnewline}{\\}
\floatstyle{ruled}
\newfloat{algorithm}{tbp}{loa}
\providecommand{\algorithmname}{Algorithm}
\floatname{algorithm}{\protect\algorithmname}

\let\oldforeign@language\foreign@language
\DeclareRobustCommand{\foreign@language}[1]{%
  \lowercase{\oldforeign@language{#1}}}

\usepackage[caption=false,font=footnotesize]{subfig}
\usepackage{amsmath}
\usepackage{cuted}
\usepackage{breqn}
\usepackage{algpseudocode}

\algblock{ParFor}{EndParFor}
\algnewcommand\algorithmicparfor{\textbf{parfor}}
\algnewcommand\algorithmicpardo{\textbf{do}}
\algnewcommand\algorithmicendparfor{\textbf{end\ parfor}}
\algrenewtext{ParFor}[1]{\algorithmicparfor\ #1\ \algorithmicpardo}
\algrenewtext{EndParFor}{\algorithmicendparfor}

\@ifundefined{showcaptionsetup}{}{%
 \PassOptionsToPackage{caption=false}{subfig}}
\usepackage{subfig}
\makeatother

\begin{document}
\title{Sensing-Assisted Channel Estimation for Flexible-Antenna Systems:
A Unified Framework}
\author{Ruoxiao~Cao,~\IEEEmembership{Graduate~Student~Member,~IEEE,}~Wentao~Yu,~\IEEEmembership{Member,~IEEE,}
Zixin~Wang,~\IEEEmembership{Member,~IEEE,}~Shenghui~Song,~\IEEEmembership{Senior~Member,~IEEE,}~Jun~Zhang,~\IEEEmembership{Fellow,~IEEE,}~\\Yi~Gong,~\IEEEmembership{Senior~Member,~IEEE,}~Khaled~B.~Letaief,~\IEEEmembership{Fellow,~IEEE}\thanks{Ruoxiao~Cao is with the Department of Electronic and Computer Engineering,
The Hong Kong University of Science and Technology, Hong Kong SAR,
and also with the Department of Electrical and Electronics Engineering,
Southern University of Science and Technology, Shenzhen 518055, China.
(e-mail: \protect\href{mailto:rcaoah@connect.ust.hk}{rcaoah@connect.ust.hk}).}\thanks{Wentao Yu is with the Department of Electrical and Computer Engineering,
The University of British Columbia, Vancouver, BC V6T 1Z4, Canada
(e-mail: \protect\href{mailto:wentaoyu@ece.ubc.ca}{wentaoyu@ece.ubc.ca}).}\thanks{Zixin~Wang, Shenghui~Song, Jun~Zhang, and Khaled~B.~Letaief are
with the the Department of Electronic and Computer Engineering, The
Hong Kong University of Science and Technology, Hong Kong SAR. (e-mail:
\protect\href{mailto:eewangzx@ust.hk}{eewangzx@ust.hk}; \protect\href{mailto:eeshsong@ust.hk}{eeshsong@ust.hk};
\protect\href{mailto:eejzhang@ust.hk}{eejzhang@ust.hk}; \protect\href{mailto:eekhaled@ust.hk}{eekhaled@ust.hk}).}\thanks{Yi~Gong is with the Department of Electrical and Electronics Engineering,
Southern University of Science and Technology, Shenzhen 518055, China
(e-mail: \protect\href{mailto:gongy@sustech.edu.cn}{gongy@sustech.edu.cn}).}}
\markboth{}{}
\maketitle
\begin{abstract}
Flexible-antenna systems, which use a small number of radio frequency
(RF) chains to dynamically access a large set of candidate antenna
locations, have emerged as a hardware-efficient architecture for 6G
networks. Acquiring accurate channel state information (CSI) is critical
for these systems, but it typically incurs a prohibitive pilot overhead
that scales with the massive number of candidate locations. To address
this bottleneck, we propose a unified sensing-assisted channel estimation
framework tailored for flexible-antenna systems. It reduces the full
CSI reconstruction problem to a consistent two-stage process: it first
resolves the dominant DOAs from the uplink data symbols by exploiting
the spatial geometry, requiring no dedicated sensing pilot, and then
calibrates the associated path gains using a minimal number of calibration
pilots. Building on this pipeline, we develop two Newton-MUSIC algorithms
tailored to different propagation environments. For line-of-sight
(LOS)-dominant environments with uncorrelated sources, we propose
SOC-Newton-MUSIC, which leverages second-order covariance (SOC) for
low-complexity DOA sensing. For non-line-of-sight (NLOS) environments
with coherent multipath, where the number of sources may exceed the
number of activated RF chains, we propose FOC-Newton-MUSIC, which
exploits fourth-order cumulants (FOC) to restore source identifiability
and structurally expand the available spatial degrees of freedom (DOFs)
through a continuous difference co-array. In both cases, by reformulating
the spatial spectrum search as a continuous optimization problem,
we replace exhaustive dense grid searches with parallelized Newton
refinements. This approach substantially reduces computational complexity
and eliminates the grid quantization errors that cause the high-SNR
performance plateau in classical grid-based MUSIC methods. Theoretical
analysis and extensive simulations confirm that the proposed framework
achieves high-resolution channel reconstruction while substantially
reducing the required pilot overhead.
\end{abstract}

\begin{IEEEkeywords}
Flexible-antenna systems, fourth-order cumulants, DOA sensing, channel
estimation, Newton refinements.
\end{IEEEkeywords}

\IEEEpeerreviewmaketitle{}

\section{Introduction}

\begin{figure}[t]
\begin{centering}
\includegraphics[scale=0.4]{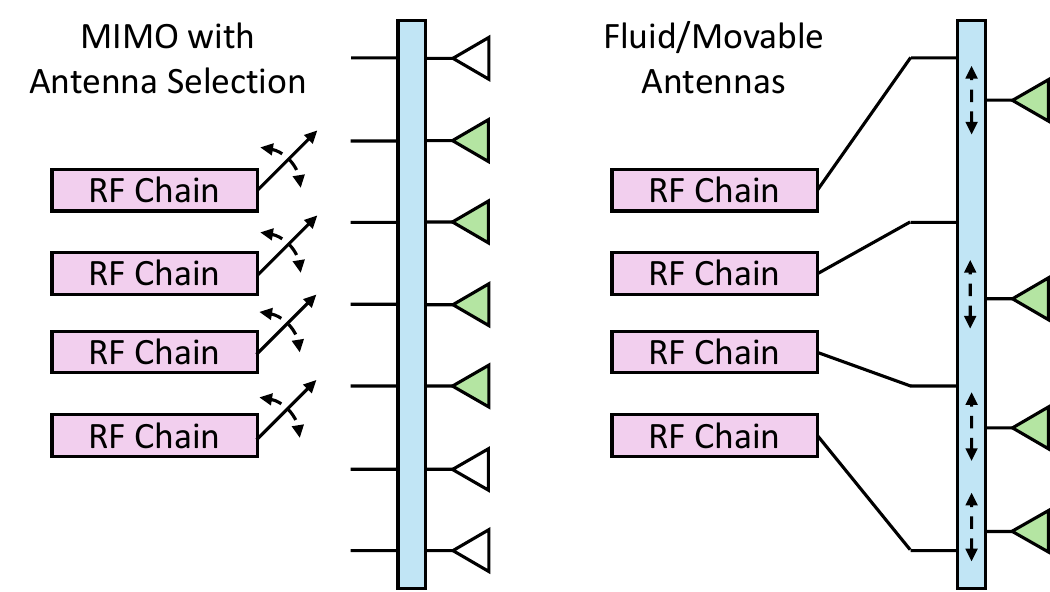}
\par\end{centering}
\caption{Two examples of flexible-antenna systems, namely MIMO systems with
antenna selection \cite{molisch2004mimo} and fluid/movable antenna
systems \cite{wong2020fluid,zhu2023movable}. \label{fig:FAS}}
\end{figure}

The evolution toward 6G networks demands unprecedented spectral efficiency,
massive connectivity, and intelligent adaptation to dynamic propagation
environments \cite{letaief2019roadmap,saad2019vision}. In this context,
flexible-antenna systems have emerged as a promising solution to intelligently
reconfigure wireless channels using limited hardware resources, leveraging
architectures with abundant switchable or reconfigurable antenna ports
but only a few activated radio frequency (RF) chains \cite{ding2025flexible}.
By decoupling the number of RF chains from the physical dimensions
of the antenna aperture, these systems offer a highly cost-effective
and energy-efficient solution. Practical realizations of this paradigm
include multiple-input multiple-output (MIMO) systems with antenna
selection \cite{molisch2004mimo}, fluid antennas \cite{wong2020fluid},
movable antennas \cite{zhu2023movable} and pinching antennas \cite{liu2025pinching},
which can dynamically synthesize diverse array geometries by selecting
optimal positions of physical elements. This capability provides spatial
degrees of freedom (DOFs) that far exceed conventional fixed-array
configurations, enabling enhanced diversity gains and superior spatial
multiplexing capabilities \cite{zhu2023movable,zhu2023movable2}.
To illustrate this paradigm, Fig. \ref{fig:FAS} depicts two typical
examples of flexible-antenna architectures, encompassing MIMO systems
with dynamic antenna selection and emerging fluid/movable antenna
configurations.

Unlocking the full potential of flexible-antenna systems relies, however,
heavily on the acquisition of accurate high-dimensional channel state
information (CSI) \cite{new2024channel}. This presents a fundamental
challenge: the CSI dimension scales proportionally with the total
number of potential antenna ports, yet only a small subset of these
ports is activated and observable at any given time. For conventional
pilot-based channel estimation schemes, estimating the full channel
matrix requires sounding all potential antenna ports exhaustively
\cite{sengijpta1995fundamentals}. Given a limited number of RF chains
and a massive number of potential ports, estimating the full channel
necessitates the base station (BS) to rapidly switch through numerous
antenna configurations, sequentially transmitting orthogonal pilot
sequences for each subset. This brute-force approach incurs prohibitive
pilot overhead and computational effort, severely reducing the time
available for actual data transmission, especially in dynamic environments
that necessitate frequent reconfigurations of the activated antenna
subset. While sparsity-based methods are often employed to alleviate
the estimation overhead, they fundamentally rely on high-dimensional
spatial codebooks associated with the massive candidate locations,
which in turn still necessitates a non-negligible amount of dedicated
pilots to identify the sparse channel supports \cite{bajwa2010compressed}.
Additionally, the scope of current sparsity-based estimation methods
in flexible-antenna systems is largely confined to single-input single-output
(SISO) configurations with a single RF chain and antenna at both the
transmitter and the receiver \cite{new2024channel,ma2023compressed}.
To address this bottleneck, sensing-assisted channel estimation has
emerged as a promising paradigm \cite{liu2022integrated}. By explicitly
exploiting the geometric properties and angular sparsity, sensing-assisted
frameworks aim to estimate the directions of arrival (DOAs), and subsequently
map these low-dimensional geometric parameters back to the high-dimensional
spatial channel, thereby bypassing the need for exhaustive port sounding
\cite{alkhateeb2014channel,el2014spatially}.

Existing sensing-assisted channel estimation strategies can generally
be classified into two distinct categories based on their signal sources:
pilot-based sensing and data payload-assisted sensing. Pilot-based
sensing methods \cite{cao2024newtonized,yu2023adaptive,cao2023joint}
dedicate reserve a specific segment of the transmission frame exclusively
for the purpose of parameter extraction. While effective, dedicating
a specialized portion of the frame to sensing inevitably reduces the
resources available for data transmission, thereby posing a challenge
for emerging systems that demand extremely high throughput. A significantly
more efficient alternative is data payload-assisted (blind) sensing
\cite{yu2026sensing,xu2024efficient,xie2025bistatic,xie2025adaptive,graff2025ofdm}.
This framework performs blind DOA sensing directly from random and
unknown uplink data symbols transmitted by the users, which completely
eliminates the need for dedicated sensing pilots. By performing DOA
sensing based on received data payloads, this strategy minimizes resource
waste and seamlessly accommodates the frequent array reconfigurations
inherent in modern flexible-antenna systems.

Regarding the implementation of this data payload-assisted sensing
strategy, existing algorithms predominantly rely on subspace-based
methods utilizing second-order covariance (SOC) \cite{krim2002two},
such as the multiple signal classification (MUSIC) algorithm \cite{schmidt1986multiple}
and the estimation of signal parameters via rotational invariance
technique (ESPRIT) \cite{roy2002esprit}. While computationally straightforward,
SOC-based techniques exhibit several theoretical limitations that
restrict their applicability in complex 6G scenarios. First, SOC inherently
assumes that all incident signal sources are statistically uncorrelated,
which is true in line-of-sight (LOS) dominant environments. However,
in non-line-of-sight (NLOS) environments with coherent multipath,
rank deficiency in the covariance matrix prevents SOC-based methods
from resolving distinct DOAs \cite{pillai1989forward}. Second, the
spatial resolution capacity of SOC is strictly upper-bounded by the
number of activated RF chains. Therefore, it fundamentally fails in
underdetermined regimes where the number of signal sources or paths
exceeds the number of available RF chains. Furthermore, unlike the
traditional uniform linear arrays (ULAs), the dynamic port selection
in flexible-antenna systems naturally yields sparse and irregular
activation patterns. In the context of such sparse linear arrays (SLAs)
\cite{li2025sparse}, classical SOC techniques often suffer from severe
grating lobes and spatial ambiguities \cite{proukakis1994study}.
Beyond these severe statistical and geometric limitations (i.e., rank
deficiency, limited DOFs, and SLA-induced grating lobes), the classical
implementation of subspace algorithms necessitates an exhaustive spatial
spectrum search over dense angular grids. This grid-based search not
only imposes a massive computational burden but also introduces unavoidable
grid quantization errors, fundamentally limiting the ultimate resolution
of DOA sensing \cite{yang2012off}.

To systematically overcome the aforementioned bottlenecks, we propose
a family of algorithms addressing both the computational burden of
exhaustive searches and the statistical limitations of SOC. First,
to eliminate the reliance on dense angular grids, we reformulate the
spatial spectrum search as a continuous optimization problem and propose
the SOC-Newton-MUSIC algorithm. By utilizing continuous Newton refinements,
this approach significantly reduces computational complexity while
mitigating grid quantization errors. Second, to address challenging
regimes in which SOC fails, including underdetermined scenarios and
NLOS environments, we propose a similar continuous optimization methodology
for the fourth-order cumulant (FOC) domain. The resulting FOC-Newton-MUSIC
framework exploits the non-Gaussian nature of practical communication
signals to synthetically expand the spatial DOFs via a continuous
difference co-array \cite{chevalier2005virtual,mendel1991tutorial}.
Finally, we show that the proposed algorithms consistently achieve
superior DOA sensing accuracy and improved channel reconstruction
performance across diverse array configurations and complex propagation
conditions.

The main contributions of this paper are summarized as follows:
\begin{itemize}
\item \emph{Sensing-Assisted Channel Estimation for Flexible-Antenna Systems}:
We develop a sensing-assisted channel estimation framework for flexible-antenna
systems that performs blind DOA sensing directly from received uplink
data symbols without dedicated sensing pilots. The framework reduces
full CSI reconstruction to a unified two-stage process: DOA estimation
from data payloads followed by path-gain calibration using a minimal
number of pilots. The full CSI is then reconstructed from the estimated
angular and path-gain parameters. This formulation decouples the pilot
overhead from the massive number of potential antenna ports $N$. 
\item \emph{SOC-Newton-MUSIC Algorithm for LOS-Dominant Scenarios}: We propose
an SOC-Newton-MUSIC algorithm for LOS-dominant scenarios, where incident
signals are uncorrelated and the number of sources is smaller than
that of activated antennas. The proposed algorithm provides substantial
savings in complexity by reformulating the spatial spectrum search
as a continuous optimization problem. Instead of exhaustively searching
over dense grids, it employs parallel Newton refinements to continually
refine coarse initial estimates into precise DOAs. Furthermore, it
effectively mitigates grid quantization errors, and breaks the estimation
performance plateau typically suffered by classical grid-based MUSIC
methods at high signal-to-noise ratios (SNRs).
\item \emph{FOC-Newton-MUSIC Framework for NLOS and Underdetermined Scenarios}:
To overcome the theoretical bottleneck of SOC in complex propagation
environments with NLOS paths and dense sources, we extend our sensing
framework to the FOC domain. By mapping the activated antenna ports
into an expanded continuous difference co-array, this approach structurally
increases the available DOFs. As a result, the framework can successfully
resolve DOAs in underdetermined scenarios, robustly restore the rank
of coherent sources in severe multipath fading, and suppress the grating
lobes that typically plague sparse antenna selections. However, evaluating
the FOC spatial spectrum by exhaustively searching over dense grids
would impose a prohibitive computational burden. To mitigate this
complexity, we extend the aforementioned continuous optimization framework
to the higher-order domain and propose an FOC-Newton-MUSIC algorithm.
\item \emph{A Unified Framework Across Diverse Scenarios}: The proposed
sensing-assisted channel estimation framework unifies the SOC and
FOC approaches within the same two-stage pipeline, namely DOA sensing
followed by path-gain calibration. Specifically, the SOC-based design
applies to LOS-dominant scenarios with uncorrelated sources, whereas
the FOC-based design applies to NLOS and underdetermined scenarios.
Extensive simulation results confirm that this unified framework ensures
high-resolution channel reconstruction with substantially reduced
pilot overhead across diverse array configurations and propagation
conditions. Moreover, when evaluated under a given pilot overhead
budget, the framework reduces the channel estimation error by a factor
of $\mathcal{O}(N/K)$ compared to conventional pilot-based methods,
where $K$ is the number of sources. 
\end{itemize}
\par The rest of this paper is organized as follows. Section \ref{sec:SystemModel}
introduces the system model and formulates the sensing-assisted channel
estimation problem tailored for flexible-antenna systems. Section
\ref{sec:Proposed-Newton-MUSIC-Algorithm} proposes the SOC-Newton-MUSIC
algorithm for LOS-dominant scenarios and discusses its theoretical
limitations. Section \ref{sec:Proposed-Fourth-Order-Newton-MUS} extends
the continuous optimization methodology to higher-order statistical
statistics and establishes the FOC-Newton-MUSIC algorithm. Section
\ref{sec:Complexity-and-Overhead} provides an analysis of computational
complexity and pilot overhead reduction. Section \ref{sec:Simulation-Results}
presents extensive simulation results to validate the effectiveness
of the proposed algorithms. Finally, Section \ref{sec:Conclusion}
concludes the paper.

\emph{Notations:} For any vector, For any matrix $\mathbf{A}$, $\mathbf{A}^{\textrm{T}}$,
$\mathbf{A}^{-1}$, $\mathbf{A}^{*}$, $\mathbf{A}^{\textrm{H}}$
and $\left[\mathbf{A}\right]_{i,j}$ denote the transpose, inverse,
conjugate, conjugate transpose and the $(i,j)$-th element of $\mathbf{A}$,
respectively. $\mathbf{I}_{N}$ represents the identity matrix of
size $N\times N$. The Euclidean norm, Frobenius norm, ceiling function
and cardinality are written as $\|\cdot\|_{2}$, $\|\cdot\|_{F}$,
$\ensuremath{\lceil\cdot\rceil}$ and $\left|\cdot\right|$, respectively.
$\mathcal{CN}(\boldsymbol{\mu},\mathbf{\Sigma})$ denotes the complex
Gaussian distribution with mean vector $\boldsymbol{\mu}$ and covariance
matrix $\mathbf{\Sigma}$. $\mathcal{U}\left(a,b\right)$ denotes
the uniform distribution from $a$ to $b$. The operators $\operatorname{diag}(\cdot)$,
$\mathbb{E}\{\cdot\}$ and $\mathfrak{R}\{\cdot\}$ generate a diagonal
matrix, calculate the expectation, and extract the real part, respectively.
The imaginary unit is $j$. $\otimes$ denotes the Kronecker product.

\section{System Model and Problem Formulation\label{sec:SystemModel}}

In this section, we first introduce the flexible-antenna architecture
and signal model. Then, we formulate the sensing-assisted channel
estimation problem tailored for flexible-antenna systems. 
\begin{figure}[t]
\begin{centering}
\includegraphics[scale=0.5]{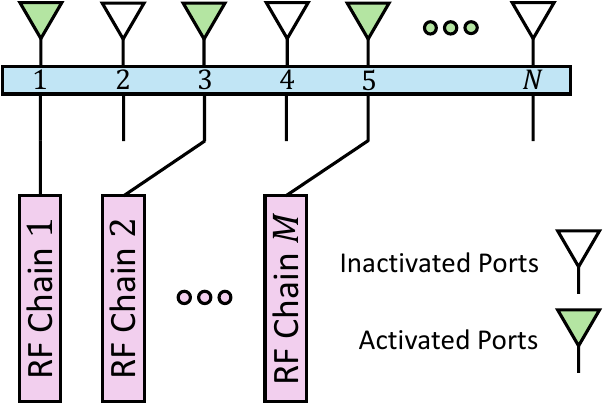}
\par\end{centering}
\caption{Flexible-antenna architecture example with $N$ potential antenna
ports and $M$ RF chains. \label{fig:architecture}}
\end{figure}

\subsection{Flexible-Antenna Architecture}

Consider the uplink of a multi-user MIMO system where the BS is equipped
with a flexible-antenna architecture. As illustrated in Fig. \ref{fig:architecture},
the system consists of $N$ potential antenna ports distributed over
a large spatial aperture. Due to hardware cost and power consumption
constraints, the system is equipped with a limited number of RF chains
$M\ll N$. At the beginning of each frame, the BS selects and activates
a subset of $M$ antenna ports to receive signals.
\begin{figure}[t]
\begin{centering}
\includegraphics[scale=0.5]{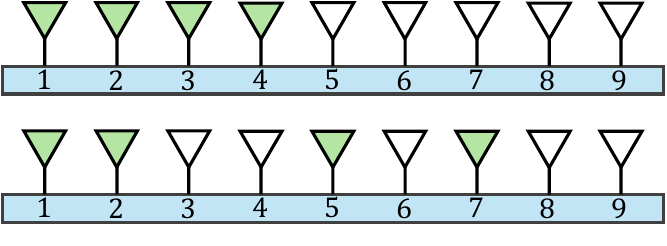}
\par\end{centering}
\caption{A comparison of different configurations for a flexible-antenna system
with $N=9$ potential antenna ports (uniform half-wavelength spacing)
and $M=4$ RF chains. The top configuration corresponds to a uniform
linear array (ULA) with $\Omega=\{1,2,3,4\}$. The bottom configuration
corresponds to a minimum redundancy array (MRA) \cite{moffet1968minimum}
with $\Omega=\{1,2,5,7\}$. \label{fig:SLAcomparison}}
\end{figure}

Let $\mathcal{N}=\{1,2,\dots,N\}$ denote the set of all available
antenna port indices. Within each coherence block, a subset of indices
$\Omega\subset\mathcal{N}$ with $|\Omega|=M$ is selected for activation.
We define an antenna selection matrix $\mathbf{\Gamma}_{\Omega}\in\{0,1\}^{M\times N}$,
where the entry $[\mathbf{\Gamma}_{\Omega}]_{m,n}=1$ if the $n$-th
port is mapped to the $m$-th RF chain, and $0$ otherwise. Accordingly,
the property $\mathbf{\Gamma}_{\Omega}\mathbf{\Gamma}_{\Omega}^{\mathsf{H}}=\mathbf{I}_{M}$
holds. Fig. \ref{fig:SLAcomparison} illustrates a comparison of two
different configurations for a flexible-antenna system with $N=9$
potential antenna ports with a uniform half-wavelength spacing and
$M=4$ RF chains. Specifically, the ULA represents the most conventional
array structure in signal processing, whereas the MRA \cite{moffet1968minimum}
is optimized to guarantee a maximum hole-free difference co-array.

\subsection{Signal Model}

Assume that $K$ narrowband far-field sources\footnote{In this paper, \textquotedbl sources\textquotedbl{} is a general
term that includes both the actual users and the coherent multipath
scatterers.} transmit simultaneously to the BS in the uplink. Let $\mathbf{s}(t)=\left[s_{1}(t),s_{2}(t),\ldots,s_{K}(t)\right]^{\mathsf{T}}\in\mathbb{C}^{K\times1}$
denote the transmitted signal vector at snapshot $t$, where $s_{k}(t)$
represents the modulated symbol of the $k$-th source. The received
signal vector at $M$ activated antenna ports $\mathbf{y}_{\Omega}(t)\in\mathbb{C}^{M\times1}$
is
\begin{equation}
\mathbf{y}_{\Omega}(t)=\mathbf{H}_{\Omega}\mathbf{s}(t)+\mathbf{n}_{\Omega}(t),\label{eq:ReceivedSignal}
\end{equation}
where $\mathbf{H}_{\Omega}\in\mathbb{C}^{M\times K}$ is the effective
channel matrix, and $\mathbf{n}_{\Omega}(t)\sim\mathcal{C}\mathcal{N}\left(\mathbf{0},\sigma^{2}\mathbf{I}_{M}\right)$
is the additive white Gaussian noise (AWGN) vector.

To capture the inherent angular sparsity of the propagation environment,
a geometric channel model is adopted. The full channel matrix $\mathbf{H}\in\mathbb{C}^{N\times K}$
corresponding to all $N$ potential ports is modeled as
\begin{equation}
\mathbf{H}=\mathbf{A}(\boldsymbol{\theta})\mathbf{P},\label{eq:ChannelModel}
\end{equation}
where $\mathbf{A}(\boldsymbol{\theta})=\left[\mathbf{a}\left(\theta_{1}\right),\ldots,\mathbf{a}\left(\theta_{K}\right)\right]\in\mathbb{C}^{N\times K}$
is the array response matrix, and $\mathbf{P}=\operatorname{diag}\left(p_{1},p_{2},\ldots,p_{K}\right)\in\mathbb{C}^{K\times K}$
is a diagonal matrix containing the complex path gains. The vector
$\boldsymbol{\theta}=\left[\theta_{1},\ldots,\theta_{K}\right]^{\mathsf{T}}$
denotes the DOAs of the sources. Denote $\mathbf{d}=[d_{1},d_{2},\dots,d_{N}]^{\mathsf{T}}\in\mathbb{R}^{N\times1}$
as the positions of potential antenna ports, the steering vector according
to DOA $\theta_{k}$ is given by
\begin{equation}
\mathbf{a}\left(\theta_{k}\right)=\left[1,e^{j2\pi\frac{d_{2}-d_{1}}{\lambda}\sin\left(\theta_{k}\right)},\ldots,e^{j2\pi\frac{d_{N}-d_{1}}{\lambda}\sin\left(\theta_{k}\right)}\right]^{\mathsf{T}},\label{eq:SteeringVector}
\end{equation}
 where $\lambda$ is the wavelength. Consequently, the effective channel
for the activated subset of antenna ports $\Omega$ is $\mathbf{H}_{\Omega}=\boldsymbol{\Gamma}_{\Omega}\mathbf{H}=\mathbf{A}_{\Omega}(\boldsymbol{\theta})\mathbf{P}$,
where $\mathbf{A}_{\Omega}(\boldsymbol{\theta})=\boldsymbol{\Gamma}_{\Omega}\mathbf{A}(\boldsymbol{\theta})$.

\subsection{Problem Formulation \label{subsec:Problem-Formulation}}

The primary objective is to estimate the high-dimensional CSI $\mathbf{H}$
with minimal pilot overhead. In flexible-antenna systems where $M\ll N$,
conventional pilot-based methods face a fundamental scalability challenge:
the pilot overhead required to estimate the channel of all potential
ports becomes prohibitive as $N$ increases. 
\begin{figure}[t]
\begin{centering}
\includegraphics[scale=0.35]{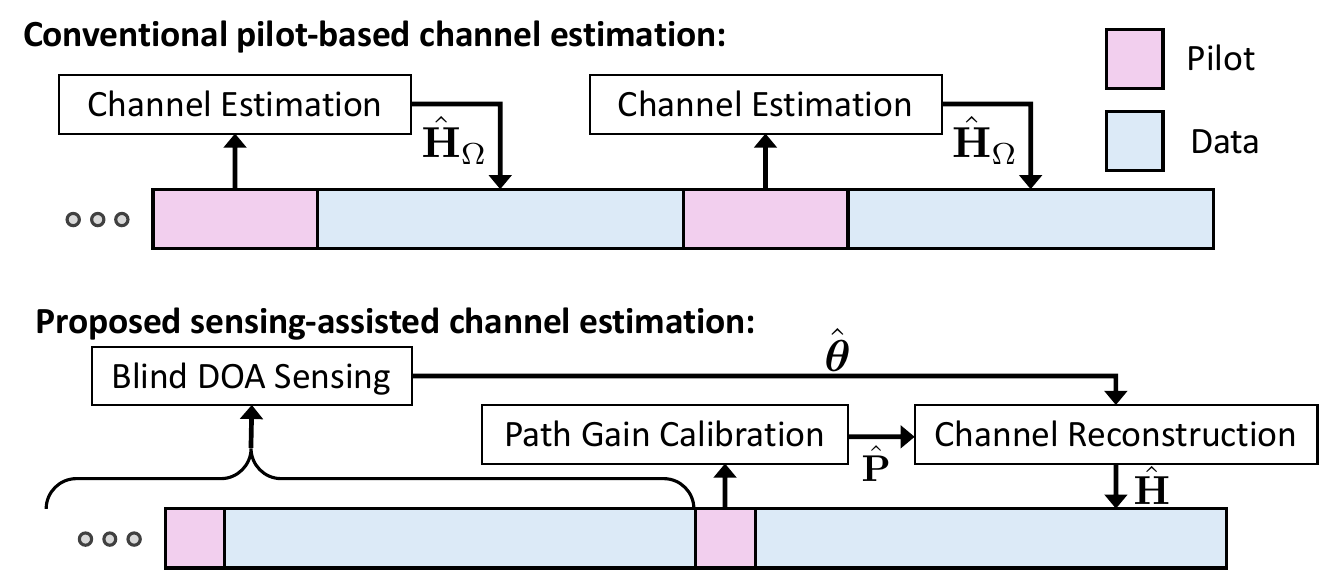}
\par\end{centering}
\caption{Comparison of the proposed sensing-assisted channel estimation with
the conventional pilot-based channel estimation. \label{fig:frame}}
\end{figure}

To address this challenge, we exploit the geometric structure of the
wireless channel to mathematically decompose the estimation task into
three cascaded sub-problems: blind DOA sensing, path gain calibration,
and channel reconstruction. Given the activated antenna subset $\Omega$
and the received data symbol matrix $\mathbf{Y}_{\Omega}=\left[\mathbf{y}_{\Omega}(1),\ldots,\mathbf{y}_{\Omega}(T)\right]\in\mathbb{C}^{M\times T}$
over $T$ snapshots, we aim to recover the channel parameters while
minimizing reliance on dedicated pilots. In particular, as shown in
Fig. \ref{fig:frame}, the sub-problems are characterized as follows:
\begin{itemize}
\item \emph{Blind DOA Sensing}: The DOA estimates $\hat{\boldsymbol{\theta}}$
are blindly extracted from the statistical properties of the received
uplink data symbols, requiring no dedicated pilot. As shown in the
frame structure, the sensing process can reuse uplink data symbols
from previous blocks to continuously refine the estimates without
incurring additional overhead. 
\item \emph{Path Gain Calibration}: Once $\hat{\boldsymbol{\theta}}$ is
acquired, the array manifold $\mathbf{A}_{\Omega}(\hat{\boldsymbol{\theta}})$
is treated as a deterministic factor. Using $T_{\textrm{p}}$ orthogonal
calibration pilots $\mathbf{\Phi}\in\mathbb{C}^{K\times T_{\textrm{p}}}$,
the calibration task is formulated as a linear least-squares (LS)
estimation problem, aimed at minimizing the residual power of the
received pilot signal $\mathbf{Y}_{\textrm{p}}\in\mathbb{C}^{M\times T_{\textrm{p}}}$:
\begin{equation}
\hat{\mathbf{P}}=\arg\min_{\mathbf{P}}\|\mathbf{Y}_{\textrm{p}}-\mathbf{A}_{\Omega}(\hat{\boldsymbol{\theta}})\mathbf{P}\mathbf{\Phi}\|_{F}^{2}.\label{eq:gainEstimation}
\end{equation}
\item \emph{Channel Reconstruction}: The full CSI is reconstructed by
\begin{equation}
\hat{\mathbf{H}}=\mathbf{A}(\hat{\boldsymbol{\theta}})\hat{\mathbf{P}}.\label{eq:channelReconstruction}
\end{equation}
\end{itemize}
Crucially, this formulation shifts the computational complexity from
high-dimensional full-port channel estimation to low-dimensional parameter
estimation. Furthermore, $\hat{\boldsymbol{\theta}}$ can be estimated
once and reused across diverse configurations $\Omega$, necessitating
only the recalibration of path gains $\hat{\mathbf{P}}$. This property
renders the framework well-suited for dynamic flexible-antenna systems
where frequent reconfiguration is required.

The performance of the reconstructed channel matrix $\hat{\mathbf{H}}$
is evaluated using the normalized mean square error (NMSE), given
by
\begin{equation}
\mathrm{NMSE}=\mathbb{E}\left\{ \frac{\|\mathbf{H}-\hat{\mathbf{H}}\|_{F}^{2}}{\|\mathbf{H}\|_{F}^{2}}\right\} .\label{eq:NMSE}
\end{equation}
For the angular parameters, the root mean square error (RMSE) of the
DOA estimates $\hat{\boldsymbol{\theta}}$ is defined as
\begin{equation}
\operatorname{RMSE}(\hat{\boldsymbol{\theta}},\boldsymbol{\theta})=\frac{1}{K}\min_{\Pi\in\mathcal{P}_{K}}\|\mathbf{\Pi}\hat{\boldsymbol{\theta}}-\boldsymbol{\theta}\|_{2},\label{eq:MSEDoA}
\end{equation}
where $\mathbf{\Pi}$ represents a specific $K\times K$ permutation
matrix while $\mathcal{P}_{K}$ denotes the set of all possible $K\times K$
permutation matrices used to resolve the ambiguity in angle ordering.

The design objectives are to minimize the channel estimation and DOA
sensing errors, as defined in \eqref{eq:NMSE} and \eqref{eq:MSEDoA},
respectively, subject to the hardware constraint $M\ll N$ and the
limited pilot budget $T_{\textrm{p}}\ll N$. However, the practical
realization of such a sensing-assisted framework must overcome several
formidable challenges. First, the algorithm must remain robust in
practical propagation environments, necessitating the ability to distinguish
coherent multipath components typically encountered in NLOS scenarios.
Second, and more critically, the framework must resolve the underdetermined
regime where the number of sources exceeds the number of activated
RF chains (i.e., $K>M$).

To systematically address these challenges, we develop the proposed
framework in a progressive manner. In Section \ref{sec:Proposed-Newton-MUSIC-Algorithm},
we first establish a performance baseline by considering a simplified
LOS-dominant scenario with fewer sources than activated antennas ($K<M$),
where signals are assumed to be uncorrelated. For this case, we propose
a computationally efficient algorithm leveraging SOC. Building upon
this, Section \ref{sec:Proposed-Fourth-Order-Newton-MUS} extends
the framework to a more general and challenging regime, characterized
by NLOS propagation with coherent multipath sources and underdetermined
conditions ($K>M$). This is accomplished through a FOC based framework,
which exploits the increased DOFs provided by the virtual array manifold
to achieve superior estimation accuracy and source identifiability.

\section{Proposed Second-Order Newton-MUSIC Algorithm \label{sec:Proposed-Newton-MUSIC-Algorithm}}

For LOS-dominant scenarios and assuming the number of sources satisfies
$K<M$, the framework leverages the computational efficiency of SOC-based
methods. These scenarios naturally align with the sparse channels
typical of millimeter-wave and terahertz communications \cite{heath2016overview,rappaport2019wireless}.
We first review the conventional SOC-MUSIC algorithm. Then, we propose
a Newton-MUSIC algorithm that enhances estimation performance while
significantly reducing computational complexity through efficient
optimization techniques.

\subsection{Review of Classical SOC-MUSIC Algorithms}

The MUSIC algorithm is the most prevalent subspace-based method for
DOA sensing. In the conventional SOC-MUSIC \cite{schmidt1986multiple},
the SOC $\hat{\mathbf{R}}\in\mathbb{C}^{M\times M}$ is estimated
from the received snapshots $\mathbf{y}_{\Omega}(t)$ as
\begin{equation}
\hat{\mathbf{R}}=\frac{1}{T}\sum_{t=1}^{T}\mathbf{y}_{\Omega}(t)\mathbf{y}_{\Omega}^{\mathsf{H}}(t).\label{eq:SCM}
\end{equation}
Performing eigenvalue decomposition (EVD) on $\hat{\mathbf{R}}$ yields
\begin{align}
\hat{\mathbf{R}} & =\mathbf{U}^{\text{(SOC)}}\mathbf{\Lambda}^{\text{(SOC)}}\left(\mathbf{U}^{\text{(SOC)}}\right)^{\mathsf{H}}\nonumber \\
 & =\mathbf{U}_{\textrm{s}}^{\text{(SOC)}}\mathbf{\Lambda}_{\textrm{s}}^{\text{(SOC)}}\left(\mathbf{U}_{\textrm{s}}^{\text{(SOC)}}\right)^{\mathsf{H}}+\mathbf{U}_{\textrm{n}}^{\text{(SOC)}}\mathbf{\Lambda}_{\textrm{n}}^{\text{(SOC)}}\left(\mathbf{U}_{\textrm{n}}^{\text{(SOC)}}\right)^{\mathsf{H}},\label{eq:EVD}
\end{align}
where $\mathbf{U}^{\text{(SOC)}}=[\mathbf{U}_{\textrm{s}}^{\text{(SOC)}},\mathbf{U}_{\textrm{n}}^{\text{(SOC)}}]$
contains all eigenvectors, $\mathbf{\Lambda}^{\text{(SOC)}}=\text{diag}(\mathbf{\Lambda}_{\textrm{s}}^{\text{(SOC)}},\mathbf{\Lambda}_{\textrm{n}}^{\text{(SOC)}})$
is the diagonal matrix of eigenvalues in descending order. The matrix
$\mathbf{U}_{\textrm{s}}^{\text{(SOC)}}\in\mathbb{C}^{M\times K}$
consists of the $K$ eigenvectors corresponding to the $K$ largest
eigenvalues that span the signal subspace, while $\mathbf{U}_{\textrm{n}}^{\text{(SOC)}}\in\mathbb{C}^{M\times(M-K)}$
consists of the remaining eigenvectors that span the noise subspace.

The basis of MUSIC is the orthogonality between the signal and noise
subspaces. Since the signal subspace is spanned by the columns of
the array response matrix $\mathbf{A}(\boldsymbol{\theta})$, any
true steering vector must be orthogonal to the noise subspace, i.e.,
$\mathbf{a}_{\Omega}^{\mathsf{H}}(\theta_{k})\mathbf{U}_{\textrm{n}}^{\text{(SOC)}}=\mathbf{0}$.
Consequently, $\left\Vert \left(\mathbf{U}_{\textrm{n}}^{\text{(SOC)}}\right)^{\mathsf{H}}\mathbf{a}_{\Omega}(\theta_{k})\right\Vert ^{2}$
vanishes when $\theta_{k}$ coincides with a true DOA. The spatial
spectrum is defined as the reciprocal of the projection of the physical
steering vector onto the noise subspace, given by
\begin{equation}
F_{\text{SOC}}(\theta)=\frac{1}{\mathbf{a}_{\Omega}^{\mathsf{H}}(\theta)\mathbf{U}_{\textrm{n}}^{\text{(SOC)}}\left(\mathbf{U}_{\textrm{n}}^{\text{(SOC)}}\right)^{\mathsf{H}}\mathbf{a}_{\Omega}(\theta)}.\label{eq:fSOC}
\end{equation}
Theoretically, $F_{\text{SOC}}(\theta)$ approaches infinity at $\theta=\theta_{k}$
creating distinct peaks that identify the true angles. To locate the
peaks of \eqref{eq:fSOC}, classical algorithms perform an exhaustive
search over discrete angular grids $\Theta$ covering the region of
interest. The estimation problem is formulated as finding the $K$
largest local maxima
\begin{equation}
\hat{\theta}_{k}=\arg\max_{\theta\in\Theta}F_{\text{SOC}}(\theta),\label{eq:optProblem}
\end{equation}
where $\hat{\boldsymbol{\theta}}=\left\{ \hat{\theta}_{k}:k=1,2,\ldots,K\right\} $
is the estimated DOAs.
\begin{algorithm}[t]
\caption{Proposed SOC-Newton-MUSIC Algorithm \label{alg:NMUSIC-SOC}}

\textbf{Input:} Received snapshots $\mathbf{Y}_{\Omega}$, number
of sources $K$, coarse grids $\Theta_{\textrm{Newton}}$, threshold
$\epsilon$, maximum number of iterations $Z$;

\begin{algorithmic}[1]

\State Compute the approximated SOC $\hat{\mathbf{R}}$ in \eqref{eq:SCM};

\State Perform EVD on $\hat{\mathbf{R}}$ to obtain $\mathbf{U}_{\textrm{n}}^{\text{(SOC)}}$
by \eqref{eq:EVD};

\State Evaluate the cost function $J_{\text{SOC}}(\theta)$ on the
coarse grids $\Theta_{\textrm{Newton}}$ and find its $K$ smallest
local minima $\hat{\boldsymbol{\theta}}^{(0)}$;

\ParFor{$k=1:K$}

\For{$z=1:Z$}

\State Compute the steering vector and its derivatives $\mathbf{a}_{\Omega}(\hat{\theta}_{k}^{(z-1)})$,
$\dot{\mathbf{a}}_{\Omega}(\hat{\theta}_{k}^{(z-1)})$, $\ddot{\mathbf{a}}_{\Omega}(\hat{\theta}_{k}^{(z-1)})$
by \eqref{eq:SteeringVector}\eqref{eq:da}\eqref{eq:dda};

\State Calculate the gradient and the Hessian of cost function $\ensuremath{\nabla J_{\text{SOC}}(\theta_{\text{curr}})}$
with \eqref{eq:dcost}\eqref{eq:ddcost};

\State Update $\hat{\theta}_{k}^{(z)}$ by the Newton's method in
\eqref{eq:newton};

\EndFor

\State $\hat{\theta}_{k}=\hat{\theta}_{k}^{(z)}$;

\EndParFor

\end{algorithmic}

\textbf{Output:} Estimated DOAs $\hat{\boldsymbol{\theta}}$.
\end{algorithm}

\subsection{Proposed Newton-MUSIC Algorithm for SOC \label{subsec:Newton-MUSIC-Algorithm-for}}

To overcome the computational limitations of the exhaustive search
over dense grids in \eqref{eq:optProblem}, we propose the Newton-MUSIC
algorithm, which reformulates the spatial spectrum search as a continuous
optimization problem. Instead of evaluating the spatial spectrum $F_{\text{SOC}}(\theta)$
over dense grids, the proposed Newton-MUSIC establishes a continuous
refinement mechanism that iteratively updates initial coarse angle
estimates, progressively driving them toward the true DOAs. This approach
significantly reduces computational complexity and eliminates off-grid
errors inherent in searching methods over discrete grids. 

The proposed Newton's method requires a reformulation of the DOA sensing
problem in \eqref{eq:optProblem}. We define the cost function $J_{\text{SOC}}(\theta)$
as
\begin{equation}
J_{\text{SOC}}(\theta)=\mathbf{a}^{\mathsf{H}}(\theta)\mathbf{U}_{\textrm{n}}^{\text{(SOC})}(\mathbf{U}_{\textrm{n}}^{\text{(SOC)}})^{\mathsf{H}}\mathbf{a}(\theta).\label{eq:costFunc}
\end{equation}
Theoretically, $J_{\text{SOC}}(\theta)=0$ when $\theta$ coincides
with a true DOA in noiseless condition. Thus, instead of searching
for the peaks of the spatial spectrum $F_{\text{SOC}}(\theta)$, the
estimation problem is formulated as finding the $K$ smallest local
minima of $J_{\text{SOC}}(\theta)$, which can be expressed by
\begin{equation}
\hat{\theta}_{k}=\arg\min_{\theta}J_{\text{SOC}}(\theta).\label{eq:optcost}
\end{equation}
We employ coarse grids $\Theta_{\textrm{Newton}}$ search with a large
spacing to identify the regions of interest. The $K$ deepest troughs
of $J_{\text{SOC}}(\theta)$ on the coarse grids serve as the initial
estimates $\hat{\boldsymbol{\theta}}^{(0)}=\left\{ \hat{\theta}_{k}^{(0)}:k=1,2,\ldots,K\right\} $.
Since the coarse search only needs to ensure the initial estimates
fall within the basin of attraction, the number of grid points is
drastically reduced compared to the exhaustive search over dense grids
in classical MUSIC algorithm. Given an initial estimate $\hat{\theta}_{k}^{(0)}$,
the Newton's method updates the estimate iteratively as
\begin{equation}
\hat{\theta}_{k}^{(z)}=\hat{\theta}_{k}^{(z-1)}-\left[\nabla^{2}J_{\text{SOC}}(\hat{\theta}_{k}^{(z-1)})\right]^{-1}\nabla J_{\text{SOC}}(\hat{\theta}_{k}^{(z-1)}),\label{eq:newton}
\end{equation}
where $\nabla J_{\text{SOC}}(\theta)$ and $\nabla^{2}J_{\text{SOC}}(\theta)$
denote the gradient and the Hessian of $J_{\text{SOC}}(\theta)$ with
respect to $\theta$, respectively. The superscript $z\in\left\{ 1,2,\ldots,Z\right\} $
denotes the iteration index. The Hessian inverse $\left[\nabla^{2}J_{\text{SOC}}(\hat{\theta}_{k}^{(z-1)})\right]^{-1}$
is simply the reciprocal of the second derivative, since $\hat{\theta}_{k}^{(z-1)}$
is a scalar parameter for the refinement stage of the $k$-th source.

To implement \eqref{eq:optcost}, we derive the analytical expressions
for the gradient and the Hessian. Denote 
\begin{equation}
\dot{\mathbf{a}}_{\Omega}(\theta)\triangleq\frac{\partial\mathbf{a}_{\Omega}(\theta)}{\partial\theta}=j\pi\cos(\theta)\mathbf{D}\mathbf{a}_{\Omega}(\theta),\label{eq:da}
\end{equation}
 
\begin{equation}
\ddot{\mathbf{a}}_{\Omega}(\theta)\triangleq\frac{\partial^{2}\mathbf{a}_{\Omega}(\theta)}{\partial\theta^{2}}=-\left(j\pi\sin(\theta)\mathbf{D}+\pi^{2}\cos^{2}(\theta)\mathbf{D}^{2}\right)\mathbf{a}_{\Omega}(\theta),\label{eq:dda}
\end{equation}
as the first and second derivatives of the steering vector, where
$\mathbf{D}\triangleq\operatorname{diag}\left(\mathbf{d}_{\Omega}\right)$
with $\mathbf{d}_{\Omega}=\boldsymbol{\Gamma}_{\Omega}\mathbf{d}$.
The gradient of the cost function is given by 
\begin{equation}
\nabla J_{\text{SOC}}(\theta)=2\Re\left\{ \dot{\mathbf{a}}_{\Omega}^{\mathsf{H}}(\theta)\mathbf{U}_{\textrm{n}}^{\text{(SOC})}(\mathbf{U}_{\textrm{n}}^{\text{(SOC)}})^{\mathsf{H}}\mathbf{a}_{\Omega}(\theta)\right\} ,\label{eq:dcost}
\end{equation}
and the Hessian is given by 
\begin{multline}
\nabla^{2}J_{\text{SOC}}(\theta)=2\Re\left\{ \dot{\mathbf{a}}_{\Omega}^{\mathsf{H}}(\theta)\mathbf{U}_{\textrm{n}}^{\text{(SOC})}(\mathbf{U}_{\textrm{n}}^{\text{(SOC)}})^{\mathsf{H}}\dot{\mathbf{a}}_{\Omega}(\theta)\right.\\
\left.+\mathbf{a}_{\Omega}^{\mathsf{H}}(\theta)\mathbf{U}_{\textrm{n}}^{\text{(SOC})}(\mathbf{U}_{\textrm{n}}^{\text{(SOC)}})^{\mathsf{H}}\ddot{\mathbf{a}}_{\Omega}(\theta)\right\} .\label{eq:ddcost}
\end{multline}

The iteration terminates after a maximum number of iterations $Z$.
The complete procedure of the proposed SOC-Newton-MUSIC algorithm
is summarized in \textbf{Algorithm} \ref{alg:NMUSIC-SOC}. Compared
to the classical SOC-MUSIC algorithm, SOC-Newton-MUSIC replaces the
exhaustive search over dense grids with Newton refinements, offering
substantial savings in computational complexity while achieving higher
estimation accuracy. Moreover, because the optimization process for
each target source is mathematically decoupled, the Newton refinements
in Step 4-11 can be fully parallelized. Such a parallel execution
architecture can further drastically reduce the time complexity in
practical implementations. 
\begin{figure}[t]
\begin{centering}
\includegraphics[scale=0.3]{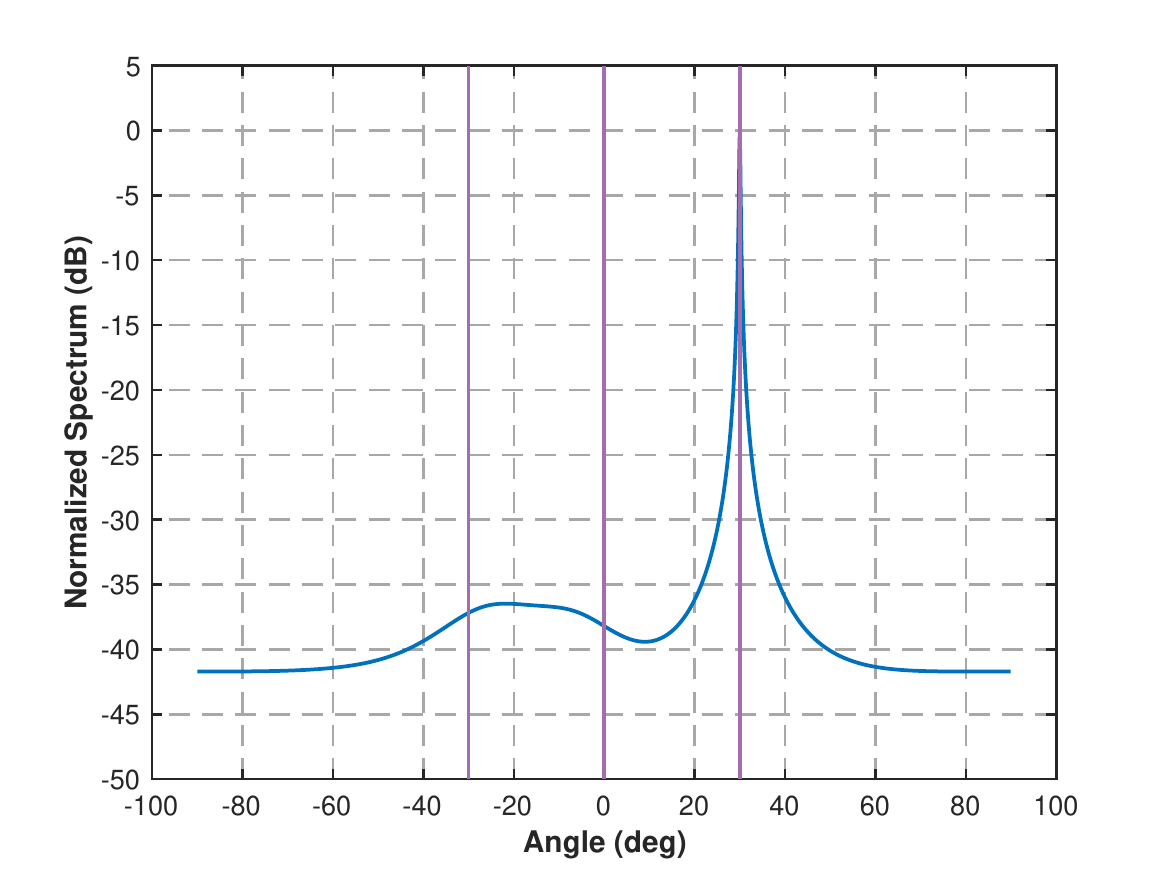}
\par\end{centering}
\caption{Normalized spatial spectrum of classical SOC-MUSIC for a $5$-antenna
ULA in the presence of coherent sources due to multipath propagation.
The sources are located at $\boldsymbol{\theta}=\{-30\lyxmathsym{\protect\textdegree},0\lyxmathsym{\protect\textdegree},30\lyxmathsym{\protect\textdegree}\}$.
The sources at $-30\lyxmathsym{\protect\textdegree}$ and $0\lyxmathsym{\protect\textdegree}$
are coherent ($s_{2}(t)=0.9s_{1}(t)$), while the source at $30\lyxmathsym{\protect\textdegree}$
is uncorrelated with the others. \label{fig:coherent_MUSIC_sectrum}}
\end{figure}

\subsection{Limitation of SOC}

While the SOC-based approaches provide a computationally efficient
baseline in LOS-dominant environments with $K<M$, it suffers from
two fundamental theoretical limitations in complex real-world propagation
scenarios.
\begin{itemize}
\item \emph{Inability to Resolve Coherent Sources}: SOC-based subspace methods
inherently rely on the assumption that incident signals are uncorrelated
(i.e., $\mathbb{E}[s_{k_{1}}(t)s_{k_{2}}^{*}(t)]=0$ for $\forall k_{1}\neq k_{2}$).
However, in severe NLOS environments characterized by multipath propagation,
multiple signals may arrive as coherent wavefronts (i.e., $s_{k_{1}}(t)=\alpha s_{k_{2}}(t)$
for distinct indices $k_{1},k_{2}$ and a complex scalar $\alpha$).
This coherence causes the source covariance matrix to become rank-deficient.
Consequently, the dimension of the signal subspace drops below $K$,
rendering SOC-based methods incapable of distinguishing the individual
DOAs. This failure is illustrative in Fig. \ref{fig:coherent_MUSIC_sectrum},
where SOC-MUSIC exhibit only a single broad plateau for two coherent
sources at $-30\lyxmathsym{\textdegree}$ and $0\lyxmathsym{\textdegree}$,
while only resolving the uncorrelated source at $30\lyxmathsym{\textdegree}$.
\item \emph{Failure in Underdetermined Regimes}: The spatial resolution
capacity of SOC is strictly upper-bounded by $M$, i.e., the number
of RF chains or activated antenna ports. Since the dimensions of the
SOC matrix are limited to $M\times M$, the signal subspace can span
at most $M-1$ dimensions. Therefore, the SOC framework fundamentally
fails in underdetermined scenarios where the number of sources exceeds
that if RF chains ($K>M$).
\end{itemize}
To address these fundamental bottlenecks, we turn to higher-order
statistical frameworks in the next section.

\section{Proposed Fourth-Order Newton-MUSIC Algorithm \label{sec:Proposed-Fourth-Order-Newton-MUS}}

Building on the framework established in Section \ref{sec:Proposed-Newton-MUSIC-Algorithm},
we extend our framework to address challenging regimes where SOC-based
methods reach their theoretical limits, specifically focusing on underdetermined
conditions ($K>M$) and NLOS environments with coherent multipath
propagation. This extension directly addresses the severe multipath
fading and massive connectivity demands (e.g., underdetermined massive
machine-type communications) prevalent in dense Sub-6GHz networks
\cite{bockelmann2016massive}. For this purpose, we establish a higher-order
statistical method based on FOC \cite{porat1991direction,peng2024under},
which offers automatic Gaussian noise suppression, virtual aperture
expansion, and rank restoration for coherent sources. Within this
framework, we first introduce the grid-search-based FOC-MUSIC. We
then generalize the Newton-MUSIC algorithm to the FOC case, demonstrating
the generality of the proposed optimization framework.

\subsection{Definition of Higher-Order Cumulants}

We first introduce the definition of cumulants via the cumulant generating
function (CGF). Let $\mathbf{x}$ be a complex random vector. The
CGF is the natural logarithm of the characteristic function, i.e.,
$\Psi_{\mathbf{x}}(\omega)=\ln\Phi_{\mathbf{x}}(\omega)$, where the
characteristic function of $\mathbf{x}$ is defined as $\Phi_{\mathbf{x}}(\omega)=\mathbb{E}\{e^{j\omega^{\mathsf{H}}\mathbf{x}}\}$.
The $q$-th order cumulant of $\mathbf{x}$ is defined as the $q$-th
order derivative of $\Psi_{\mathbf{x}}(\omega)$ evaluated at $\omega=0$.

A prominent motivation for employing higher-order cumulants (HOC)
in the proposed framework is their inherent capability for Gaussian
noise suppression. Consider a zero-mean circular complex Gaussian
noise variable $\xi\sim\mathcal{C}\mathcal{N}\left(0,\sigma^{2}\right)$.
Its CGF is $\Psi_{\xi}(\omega)=-\frac{1}{2}\sigma^{2}|\omega|^{2}$,
which is a polynomial of order $2$. According to the definition,
the $q$-th order cumulant of a Gaussian process vanishes for all
$q\geqslant3$ \cite{mendel1991tutorial}. In contrast, practical
modulated signals (e.g., QPSK, 16QAM) are non-Gaussian and possess
non-zero HOC.

\subsection{FOC Matrix Construction}

Among HOC-based methods, FOC offers a favorable balance between estimation
accuracy and computational complexity \cite{mendel1991tutorial,swami2002hierarchical},
and is therefore adopted in this work. For the received signal vector
$\mathbf{y}_{\Omega}(t)$, the FOC is defined as
\begin{multline}
\operatorname{cum}\left\{ y_{\Omega,m_{1}}(t),y_{\Omega,m_{2}}^{*}\left(t+l_{1}\right),\right.\\
\left.y_{\Omega,m_{3}}\left(t+l_{2}\right),y_{\Omega,m_{4}}^{*}\left(t+l_{3}\right)\right\} \label{eq:FOCDef}
\end{multline}
where $m_{u}\in\Omega\ (u=1,2,3,4)$ denotes the indices of the activated
antenna elements, $y_{\Omega,m_{u}}(t)$ denotes $m_{u}$-th element
of $\mathbf{y}_{\Omega}(t)$, and $l_{1}$, $l_{2}$ and $l_{3}$
are the time lags. Following standard conventions in array direction
finding \cite{porat1991direction}, we focus our analysis on the spatial
FOC by setting the time lags to zero, i.e., $l_{1}=l_{2}=l_{3}=0$,
given by 
\begin{equation}
\begin{aligned}\operatorname{cum} & \left\{ y_{\Omega,m_{1}}(t),y_{\Omega,m_{2}}^{*}(t),y_{\Omega,m_{3}}(t),y_{\Omega,m_{4}}^{*}(t)\right\} \\
= & \mathbb{E}\left\{ y_{\Omega,m_{1}}(t)y_{\Omega,m_{2}}^{*}(t)y_{\Omega,m_{3}}(t)y_{\Omega,m_{4}}^{*}(t)\right\} \\
 & -\mathbb{E}\left\{ y_{\Omega,m_{1}}(t)y_{\Omega,m_{2}}^{*}(t)\right\} \mathbb{E}\left\{ y_{\Omega,m_{3}}(t)y_{\Omega,m_{4}}^{*}(t)\right\} \\
 & -\mathbb{E}\left\{ y_{\Omega,m_{1}}(t)y_{\Omega,m_{3}}(t)\right\} \mathbb{E}\left\{ y_{\Omega,m_{2}}^{*}(t)y_{\Omega,m_{4}}^{*}(t)\right\} \\
 & -\mathbb{E}\left\{ y_{\Omega,m_{1}}(t)y_{\Omega,m_{4}}^{*}(t)\right\} \mathbb{E}\left\{ y_{\Omega,m_{2}}^{*}(t)y_{\Omega,m_{3}}(t)\right\} 
\end{aligned}
.\label{eq:FOCDef2}
\end{equation}
To exploit the non-Gaussianity of the source signals, we define the
FOC matrix $\mathbf{C}_{4}\in\mathbb{C}^{M^{2}\times M^{2}}$ with
elements
\begin{multline}
\left[\mathbf{C}_{4}\right]_{(m_{1}-1)M+m_{2},(m_{3}-1)M+m_{4}}\\
=\operatorname{cum}\left\{ y_{\Omega,m_{1}}(t),y_{\Omega,m_{2}}^{*}(t),y_{\Omega,m_{3}}(t),y_{\Omega,m_{4}}^{*}(t)\right\} .\label{eq:C4element}
\end{multline}
Substituting the signal model \eqref{eq:ReceivedSignal} into the
cumulant definition \eqref{eq:C4element} yields the following matrix-form
decomposition\footnote{It is important to note that the Kronecker product in Eq. \eqref{eq:C4Full}
strictly operates on the received signal vector $\mathbf{y}_{\Omega}(t)$
at each snapshot, rather than the data matrix $\mathbf{Y}_{\Omega}$.
The key lies in exploiting the spatial correlations among different
antenna elements at the same snapshot. However, we observe that some
existing works mistakenly construct it based on $\mathbf{Y}_{\Omega}$.
Although this does not alter the dimension of the FOC matrix, it contradicts
the physical interpretation by incorporating correlations across different
snapshots for all antenna elements, ultimately resulting in estimation
biases.}
\begin{equation}
\begin{aligned}\mathbf{C}_{4}= & \mathbb{E}\left\{ \left(\mathbf{y}_{\Omega}(t)\otimes\mathbf{y}_{\Omega}^{*}(t)\right)\left(\mathbf{y}_{\Omega}(t)\otimes\mathbf{y}_{\Omega}^{*}(t)\right)^{\mathsf{H}}\right\} \\
 & -\mathbb{E}\left\{ \mathbf{y}_{\Omega}(t)\otimes\mathbf{y}_{\Omega}^{*}(t)\right\} \mathbb{E}\left\{ \left(\mathbf{y}_{\Omega}(t)\otimes\mathbf{y}_{\Omega}^{*}(t)\right)^{\mathsf{H}}\right\} \\
 & -\mathbb{E}\left\{ \mathbf{y}_{\Omega}(t)\mathbf{y}_{\Omega}^{\mathsf{H}}(t)\right\} \otimes\mathbb{E}\left\{ \mathbf{y}_{\Omega}(t)\mathbf{y}_{\Omega}^{\mathsf{H}}(t)\right\} 
\end{aligned}
.\label{eq:C4Full}
\end{equation}
This can be compactly represented as
\begin{equation}
\mathbf{C}_{4}=\left(\mathbf{A}_{\Omega}(\boldsymbol{\theta})\otimes\mathbf{A}_{\Omega}^{*}(\boldsymbol{\theta})\right)\mathbf{C}_{\textrm{s}}\left(\mathbf{A}_{\Omega}(\boldsymbol{\theta})\otimes\mathbf{A}_{\Omega}^{*}(\boldsymbol{\theta})\right)^{\mathsf{H}},\label{eq:C4}
\end{equation}
where $\mathbf{C}_{\textrm{s}}\in\mathbb{C}^{K^{2}\times K^{2}}$
is a diagonal matrix containing the FOC of the source signals, i.e.,
\begin{equation}
\begin{aligned}\mathbf{C}_{\textrm{s}}\triangleq & \mathbb{E}\left\{ \left(\mathbf{s}(t)\otimes\mathbf{s}^{*}(t)\right)\left(\mathbf{s}(t)\otimes\mathbf{s}^{*}(t)\right)^{\mathsf{H}}\right\} \\
 & -\mathbb{E}\left\{ \mathbf{s}(t)\otimes\mathbf{s}^{*}(t)\right\} \mathbb{E}\left\{ \left(\mathbf{s}(t)\otimes\mathbf{s}^{*}(t)\right)^{\mathsf{H}}\right\} \\
 & -\mathbb{E}\left\{ \mathbf{s}(t)\mathbf{s}^{\mathsf{H}}(t)\right\} \otimes\mathbb{E}\left\{ \mathbf{s}(t)\mathbf{s}^{\mathsf{H}}(t)\right\} 
\end{aligned}
.\label{eq:Cs}
\end{equation}

\subsection{Virtual Array Interpretation and DOF Analysis}

The matrix decomposition in \eqref{eq:C4} highlights a pivotal advantage
of the FOC-based framework: the synthetic expansion of the array aperture.
This expansion significantly increases the DOFs, enabling the resolution
of more sources than the number of RF chains (i.e., $K>M$), thereby
effectively addressing the underdetermined scenarios in flexible-antenna
systems. Detailed derivations regarding the specific DOF gain are
provided in the subsequent analysis.

As observed from \eqref{eq:C4}, the FOC matrix $\mathbf{C}_{4}$
exhibits a structural analogy to the SOC matrix $\mathbf{R}$. In
this higher-order framework, the original array response matrix $\mathbf{A}_{\Omega}(\boldsymbol{\theta})$
is replaced by the Khatri-Rao product $\mathbf{A}_{\Omega}(\boldsymbol{\theta})\otimes\mathbf{A}_{\Omega}^{*}(\boldsymbol{\theta})$.
Consequently, we define the virtual steering vector as 
\begin{figure}[t]
\begin{centering}
\includegraphics[scale=0.5]{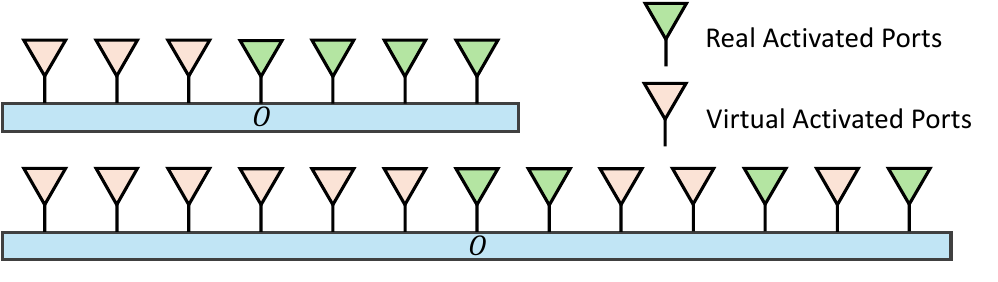}
\par\end{centering}
\caption{The virtual array of different configurations in Fig. \ref{fig:SLAcomparison},
where $O$ is the original position of the $1$-st antenna. The top
array is the difference co-array by the ULA with $\nu_{\textrm{ULA}}=6$.
The bottom array is the difference co-array by the MRA with $\nu_{\textrm{MRA}}=12$.
\label{fig:VirtualArray}}
\end{figure}
\begin{equation}
\mathbf{b}\left(\theta_{k}\right)=\mathbf{a}_{\Omega}\left(\theta_{k}\right)\otimes\mathbf{a}_{\Omega}^{*}\left(\theta_{k}\right)\in\mathbb{C}^{M^{2}\times1}.\label{eq:VirtualSteeringVector}
\end{equation}
To elucidate the geometry of this virtual array, consider the phase
term corresponding to antenna indices $(m_{1},m_{2})$ as
\begin{equation}
\left[\mathbf{b}\left(\theta_{k}\right)\right]_{(m_{1}-1)M+m_{2}}=e^{\mathrm{j}\pi\left(d_{m_{1}}-d_{m_{2}}\right)\sin\left(\theta_{k}\right)}.\label{eq:virtualPhaseTerm}
\end{equation}
It indicates that the virtual array contains elements at positions
corresponding to the \emph{difference} of activated antenna positions,
i.e., the difference co-array defined by the set $\mathcal{D}=\left\{ d_{m_{1}}-d_{m_{2}}\mid m_{1},m_{2}\in\Omega\right\} $.
Notably, although the FOC matrix $\mathbf{C}_{4}$ is of dimension
$M^{2}\times M^{2}$, the maximum number of resolvable sources, denoted
as $\nu$, is fundamentally constrained by the cardinality of the
difference co-array $\mathcal{D}$, i.e., $\nu=|\mathcal{D}|-1$,
rather than the matrix dimension $M^{2}$. This constraint defines
the DOF limit of the proposed FOC framework in underdetermined scenarios.
As illustrated by the top two arrays in Fig. \ref{fig:VirtualArray},
the virtual steering vectors are calculated by
\[
\begin{aligned}\mathbf{b}_{\textrm{ULA}}\left(\theta_{k}\right)= & \left[1,e^{j1\pi\sin(\theta)},e^{j2\pi\sin(\theta)},e^{j3\pi\sin(\theta)}\right.\\
 & e^{-j1\pi\sin(\theta)},1,e^{j1\pi\sin(\theta)},e^{j2\pi\sin(\theta)},\\
 & e^{-j2\pi\sin(\theta)},e^{-j1\pi\sin(\theta)},1,e^{j1\pi\sin(\theta)},\\
 & \left.e^{-j3\pi0\sin(\theta)},e^{-j2\pi\sin(\theta)},e^{-j1\pi\sin(\theta)},1\right]^{\mathsf{T}}
\end{aligned}
,
\]
\[
\begin{aligned}\mathbf{b}_{\textrm{MRA}}\left(\theta_{k}\right)= & \left[1,e^{j1\pi\sin(\theta)},e^{j4\pi\sin(\theta)},e^{j6\pi\sin(\theta)}\right.\\
 & e^{-j1\pi\sin(\theta)},1,e^{j3\pi\sin(\theta)},e^{j5\pi\sin(\theta)},\\
 & e^{-j4\pi\sin(\theta)},e^{-j3\pi\sin(\theta)},1,e^{j2\pi\sin(\theta)},\\
 & \left.e^{-j6\pi0\sin(\theta)},e^{-j5\pi\sin(\theta)},e^{-j2\pi\sin(\theta)},1\right]^{\mathsf{T}}
\end{aligned}
.
\]
By counting the unique elements in the virtual steering vectors, we
can see that the DOFs yielded by the ULA and the MRA are $\nu_{\textrm{ULA}}=6$
and $\nu_{\textrm{MRA}}=12$, respectively.

This difference co-array structure distinguishes our framework from
existing FOC-based methods that rely on sum co-arrays \cite{peng2024under},
where the virtual steering vector is given by $\mathbf{\mathbf{a}}_{\Omega}(\theta)\otimes\mathbf{a}_{\Omega}(\theta)$
and the array positions are defined by the sum co-array $\mathcal{S}=\left\{ d_{m_{1}}+d_{m_{2}}\mid m_{1},m_{2}\in\Omega\right\} $.
While sum co-arrays can expand the aperture, they often suffer from
holes (missing lags) and grating lobes unless the physical array is
meticulously designed. In contrast, the difference co-array generated
by \eqref{eq:VirtualSteeringVector} ensures a continuous virtual
aperture for flexible-antenna selections, effectively avoiding holes
and preventing grating lobes. 

\subsection{Classical MUSIC Algorithm for FOC}

With $T$ snapshots, $\mathbf{C}_{4}$ in \eqref{eq:C4Full} is approximated
by 
\begin{equation}
\begin{aligned}\mathbf{\hat{C}}_{4}= & \frac{1}{T}\sum_{t=1}^{T}\left\{ \left(\mathbf{y}_{\Omega}(t)\otimes\mathbf{y}_{\Omega}^{*}(t)\right)\left(\mathbf{y}_{\Omega}(t)\otimes\mathbf{y}_{\Omega}^{*}(t)\right)^{\mathsf{H}}\right\} \\
 & -\frac{1}{T}\sum_{t=1}^{T}\left\{ \mathbf{y}_{\Omega}(t)\otimes\mathbf{y}_{\Omega}^{*}(t)\right\} \frac{1}{T}\sum_{t=1}^{T}\left\{ \left(\mathbf{y}_{\Omega}(t)\otimes\mathbf{y}_{\Omega}^{*}(t)\right)^{\mathsf{H}}\right\} \\
 & -\frac{1}{T}\sum_{t=1}^{T}\left\{ \mathbf{y}_{\Omega}(t)\mathbf{y}_{\Omega}^{\mathsf{H}}(t)\right\} \otimes\frac{1}{T}\sum_{t=1}^{T}\left\{ \mathbf{y}_{\Omega}(t)\mathbf{y}_{\Omega}^{\mathsf{H}}(t)\right\} 
\end{aligned}
.\label{eq:C4hat}
\end{equation}
To separate the signal and the noise subspaces, we perform singular
value decomposition (SVD) on $\mathbf{\hat{C}}_{4}\in\mathbb{C}^{M^{2}\times M^{2}}$,
i.e.,
\begin{align}
\hat{\mathbf{C}}_{4} & =\mathbf{U}^{\text{(FOC)}}\mathbf{\Sigma^{\text{(FOC)}}}\left(\mathbf{V}^{\text{(FOC)}}\right)^{\mathsf{H}}\nonumber \\
 & =\begin{bmatrix}\mathbf{U}_{\textrm{s}}^{\text{(FOC)}} & \mathbf{U}_{\textrm{n}}^{\text{(FOC)}}\end{bmatrix}\begin{bmatrix}\mathbf{\Sigma}_{\textrm{s}}^{\text{(FOC)}} & \mathbf{0}\\
\mathbf{0} & \mathbf{\Sigma}_{\textrm{n}}^{\text{(FOC)}}
\end{bmatrix}\begin{bmatrix}\left(\mathbf{V}_{\textrm{s}}^{\text{(FOC)}}\right)^{\mathsf{H}}\\
\left(\mathbf{V}_{\textrm{n}}^{\text{(FOC)}}\right)^{\mathsf{H}}
\end{bmatrix},\label{eq:SVD}
\end{align}
where $\mathbf{U}_{\textrm{s}}^{\text{(FOC)}}\in\mathbb{C}^{M^{2}\times K}$
and $\mathbf{V}_{\textrm{s}}^{\text{(FOC)}}\in\mathbb{C}^{M^{2}\times K}$
contain the left and the right singular vectors corresponding to the
$K$ largest singular values, which span the signal subspace. The
matrices $\mathbf{U}_{\textrm{n}}^{\text{(FOC)}}\in\mathbb{C}^{M^{2}\times(M^{2}-K)}$
and $\mathbf{V}_{\textrm{n}}^{\text{(FOC)}}\in\mathbb{C}^{M^{2}\times(M^{2}-K)}$
contain the remaining singular vectors, spanning the noise subspace.
The diagonal matrices $\mathbf{\Sigma}_{\textrm{s}}^{\text{(FOC)}}$
and $\mathbf{\Sigma}_{\textrm{n}}^{\text{(FOC)}}$ contain the corresponding
singular values.
\begin{algorithm}[t]
\caption{Classical FOC-MUSIC Algorithm \label{alg:classicMUSIC}}

\textbf{Input:} Received snapshots $\mathbf{Y}_{\Omega}$, number
of sources $K$, search grids $\Theta$;

\begin{algorithmic}[1]

\State Compute the approximated FOC matrix $\mathbf{\hat{C}}_{4}$
in \eqref{eq:C4hat};

\State Perform SVD on $\mathbf{\hat{C}}_{4}$ to obtain $\mathbf{U}_{\textrm{n}}^{\text{(FOC)}}$
by \eqref{eq:SVD};

\State Compute the virtual steering vector of virtual array $\mathbf{b}\left(\theta_{k}\right)$
on the search grids $\Theta$ by \eqref{eq:VirtualSteeringVector};

\State Evaluate $F_{\text{FOC}}(\theta)$ in \eqref{eq:fFOC} on
the grids $\Theta$ and find its $K$ largest discrete local maxima,
denoted $\hat{\boldsymbol{\theta}}$.

\end{algorithmic}

\textbf{Output:} Estimated DOAs $\hat{\boldsymbol{\theta}}$.
\end{algorithm}

Similar to the SOC-MUSIC algorithm, the spatial spectrum for FOC-MUSIC
is then constructed as
\begin{equation}
F_{\text{FOC}}(\theta)=\frac{1}{\mathbf{b}^{\mathsf{H}}(\theta)\mathbf{U}_{\textrm{n}}^{\text{(FOC)}}(\mathbf{U}_{\textrm{n}}^{\text{(FOC)}})^{\mathsf{H}}\mathbf{b}(\theta)}.\label{eq:fFOC}
\end{equation}
The DOA estimates are obtained by finding the $K$ highest peaks of
$F_{\text{FOC}}(\theta)$. To locate these peaks, classical implementations
perform an exhaustive search over discretized angular grids $\Theta$.
The estimation problem is formulated as 
\begin{equation}
\hat{\theta}_{k}=\arg\max_{\theta\in\Theta}F_{\text{FOC}}(\theta).\label{eq:optproblem2}
\end{equation}
 The complete procedure is summarized in \textbf{Algorithm} \ref{alg:classicMUSIC}.

\subsection{Proposed Newton-MUSIC Algorithm for FOC}

Building upon the continuous optimization framework established in
Section \ref{subsec:Newton-MUSIC-Algorithm-for}, we now extend the
Newton-MUSIC algorithm to the higher-order case. The core distinction
lies in substituting the activated steering vector $\mathbf{a}_{\Omega}(\theta)$
with the extended virtual steering vector $\mathbf{b}(\theta)$. Consequently,
the objective function becomes
\begin{equation}
J_{\text{FOC}}(\theta)=\mathbf{b}^{\mathsf{H}}(\theta)\mathbf{U}_{\textrm{n}}^{\text{(FOC)}}(\mathbf{U}_{\textrm{n}}^{\text{(FOC)}})^{\mathsf{\mathsf{H}}}\mathbf{b}(\theta).\label{eq:costFunc-1}
\end{equation}
The estimation problem is formulated as finding the local minima
\begin{equation}
\hat{\theta}=\arg\min_{\theta}J_{\text{FOC}}(\theta).\label{eq:optcost-1}
\end{equation}
The initial estimates $\hat{\boldsymbol{\theta}}^{(0)}$ are obtained
from the coarse grids $\Theta_{\textrm{Newton}}$. By employing the
following Newton update rule iteratively
\begin{equation}
\hat{\theta}_{k}^{(z)}=\hat{\theta}_{k}^{(z-1)}-\left[\nabla^{2}J_{\text{FOC}}(\hat{\theta}_{k}^{(z-1)})\right]^{-1}\nabla J_{\text{FOC}}(\hat{\theta}_{k}^{(z-1)}),\label{eq:newton-1}
\end{equation}
we efficiently locate the DOAs without relying on exhaustive search
over dense grids . 
\begin{algorithm}[t]
\caption{Proposed FOC-Newton-MUSIC Algorithm \label{alg:NMUSIC-1}}

\textbf{Input:} Received snapshots $\mathbf{Y}_{\Omega}$, number
of sources $K$, coarse grids $\Theta_{\textrm{Newton}}$, threshold
$\epsilon$, maximum number of iterations $Z$;

\begin{algorithmic}[1]

\State Compute the approximated FOC $\mathbf{\hat{C}}_{4}$ in \eqref{eq:C4hat};

\State Perform SVD on $\mathbf{\hat{C}}_{4}$ to obtain $\mathbf{U}_{\textrm{n}}^{\text{(FOC)}}$
by \eqref{eq:SVD};

\State Evaluate the cost function $J_{\text{FOC}}(\theta)$ on the
coarse grids $\Theta_{\textrm{Newton}}$ and find its $K$ smallest
local minima $\hat{\boldsymbol{\theta}}^{(0)}$ ;

\ParFor{$k=1:K$}

\For{$z=1:Z$}

\State Compute the virtual steering vector and its derivatives $\mathbf{b}_{\Omega}(\hat{\theta}_{k}^{(z-1)})$,
$\dot{\mathbf{b}}_{\Omega}(\hat{\theta}_{k}^{(z-1)})$, $\ddot{\mathbf{b}}_{\Omega}(\hat{\theta}_{k}^{(z-1)})$
by \eqref{eq:VirtualSteeringVector}\eqref{eq:db}\eqref{eq:ddb};

\State Calculate the gradient and the Hessian of cost function $\ensuremath{\nabla J_{\text{FOC}}(\theta_{\text{curr}})}$
with \eqref{eq:dcost-1}\eqref{eq:ddcost-1};

\State Update $\hat{\theta}_{k}^{(z)}$ by the Newton's method in
\eqref{eq:newton-1};

\EndFor

\State $\hat{\theta}_{k}=\hat{\theta}_{k}^{(z)}$;

\EndParFor

\end{algorithmic}

\textbf{Output:} Estimated DOAs $\hat{\boldsymbol{\theta}}$.
\end{algorithm}

The analytical gradient and Hessian required for the update are derived
using the virtual array's specific geometry, where the first and the
second derivatives $\dot{\mathbf{b}}(\theta)$ and $\ddot{\mathbf{b}}(\theta)$
are derived as
\begin{equation}
\dot{\mathbf{b}}(\theta)=\frac{\partial\mathbf{b}(\theta)}{\partial\theta}=j\pi\cos(\theta)\mathbf{D}_{v}\mathbf{b}(\theta),\label{eq:db}
\end{equation}
 
\begin{equation}
\ddot{\mathbf{b}}(\theta)=\frac{\partial^{2}\mathbf{b}(\theta)}{\partial\theta^{2}}=-\left(j\pi\sin(\theta)\mathbf{D}_{v}+\pi^{2}\cos^{2}(\theta)\mathbf{D}_{v}^{2}\right)\mathbf{b}(\theta),\label{eq:ddb}
\end{equation}
where $\mathbf{d}_{v}=[d_{v,1},d_{v,2},\dots,d_{v,M^{2}}]^{\mathsf{T}}\in\mathbb{R}^{M^{2}\times1}$
contains the positions of the virtual array elements, and $\mathbf{D}_{v}\triangleq\operatorname{diag}(\mathbf{d}_{v})$.
The gradient and the Hessian are then given by
\begin{equation}
\nabla J_{\text{FOC}}(\theta)=2\Re\left\{ \dot{\mathbf{b}}^{\mathsf{H}}(\theta)\mathbf{\Pi}_{n}\mathbf{b}(\theta)\right\} ,\label{eq:dcost-1}
\end{equation}
\begin{equation}
\nabla^{2}J_{\text{FOC}}(\theta)=2\Re\left\{ \dot{\mathbf{b}}^{\mathsf{H}}(\theta)\mathbf{\Pi}_{n}\dot{\mathbf{b}}(\theta)+\mathbf{b}^{\mathsf{H}}(\theta)\mathbf{\Pi}_{n}\ddot{\mathbf{b}}(\theta)\right\} .\label{eq:ddcost-1}
\end{equation}
The complete procedure is summarized in \textbf{Algorithm} \ref{alg:NMUSIC-1}.
This unified approach ensures that the superior noise suppression
and virtual aperture expansion of FOC can be exploited without the
prohibitive computational burden of an exhaustive search over dense
grids.
\begin{table*}[tbh]
\caption{Complexity Comparison \label{tab:CCC}}

\centering{}%
\begin{tabular}{|l|c|c|c|c|}
\hline 
\textbf{Algorithm} & \textbf{SOC-MUSIC} & \textbf{SOC-Newton-MUSIC} & \textbf{FOC-MUSIC} & \textbf{FOC-Newton-MUSIC}\tabularnewline
\hline 
Matrix Construction & $\mathcal{O}(M^{2}T)$ & $\mathcal{O}(M^{2}T)$ & $\mathcal{O}(M^{4}T)$ & $\mathcal{O}(M^{4}T)$\tabularnewline
Subspace Decomposition & $\mathcal{O}(M^{3})$ & $\mathcal{O}(M^{3})$ & $\mathcal{O}(M^{6})$ & $\mathcal{O}(M^{6})$\tabularnewline
DOA Extraction & $\mathcal{O}(|\Theta|M)$ & $\mathcal{O}((|\Theta_{\textrm{Newton}}|+Z)M)$ & $\mathcal{O}(|\Theta|M^{2})$ & $\mathcal{O}((|\Theta_{\textrm{Newton}}|+Z)M^{2})$\tabularnewline
\hline 
\end{tabular}
\end{table*}

\section{Complexity and Overhead Analysis \label{sec:Complexity-and-Overhead}}

In this section, we analyze the complexity and the pilot overhead
of the proposed unified framework. We first examine the computational
complexity of both SOC-based and FOC-based implementations, comparing
the classical grid-search-based MUSIC algorithms with the proposed
Newton-MUSIC variants. Subsequently, we quantify the pilot overhead
reduction achieved by the sensing-assisted channel estimation compared
to conventional methods. Together, these analyses demonstrate the
efficiency gains that make the proposed framework particularly suitable
for flexible-antenna systems.

\subsection{Complexity Analysis}

The overall complexity can be decomposed into three main stages: 1)
matrix construction, involving the calculation of either the SOC or
FOC matrices; 2) subspace decomposition, performed via EVD for SOC
or SVD for FOC; and 3) DOA extraction, which encompasses the grid
search and/or subsequent optimization. Table \ref{tab:CCC} summarizes
the complexity of the four methods. These complexity differences are
mainly driven by two key transitions: the shift from second-order
to fourth-order statistics, and the replacement of exhaustive search
over dense grids with Newton refinement.

\subsubsection{Statistical Order Transition}

As shown in Table \ref{tab:CCC}, transitioning from SOC to FOC entails
a significant expansion in matrix dimensions, specifically from $M\times M$
to $M^{2}\times M^{2}$. Consequently, the computational complexity
associated with matrix construction scales from $\mathcal{O}(M^{2}T)$
to $\mathcal{O}(M^{4}T)$. The estimation accuracy of FOC-based algorithm
in the finite-snapshot regime is inherently subject to higher statistical
variance than SOC-based techniques. This necessitates a substantially
larger $T$ to facilitate the statistical convergence of HOC, which
is essential for ensuring estimation stability and accuracy \cite{porat1991direction},
which increases the computational burden in practice. Furthermore,
the complexity of the subsequent subspace decomposition scales from
$\mathcal{O}(M^{3})$ to $\mathcal{O}(M^{6})$. However, it is important
to emphasize that this complexity increase is inherent to the transition
into higher-order statistical domains and imposes an identical overhead
on both the classical MUSIC baselines and our proposed Newton-MUSIC
algorithm.

\subsubsection{DOA Extraction Paradigm}

The most substantial computational efficiency gains in our proposed
unified framework emerge during the spatial spectrum search stage.
Conventional MUSIC algorithms rely on evaluating the steering vector
over dense angular grids $\Theta$, resulting in search complexities
of $\mathcal{O}(|\Theta|M)$ for SOC and $\mathcal{O}(|\Theta|M^{2})$
for FOC. By introducing the Newton-MUSIC variants, we restrict the
spatial spectrum search to much more coarse grids $\Theta_{\textrm{Newton}}$
that only serve as initialization. The computational complexity is
offloaded to $Z$ Newton iterations, reducing the complexities to
$\mathcal{O}((|\Theta_{\textrm{Newton}}|+KZ)M)$ and $\mathcal{O}((|\Theta_{\textrm{Newton}}|+KZ)M^{2})$
for the SOC and the FOC cases, respectively. Furthermore, since each
DOA is optimized independently within the Newton-MUSIC framework,
the iterations for all $K$ sources can be executed in parallel, which
further reduces the complexities to $\mathcal{O}((|\Theta_{\textrm{Newton}}|+Z)M)$
and $\mathcal{O}((|\Theta_{\textrm{Newton}}|+Z)M^{2})$ for the SOC
and the FOC cases, respectively. Given that $|\Theta_{\textrm{Newton}}|\ll|\Theta|$
and $KZ\ll|\Theta|$, the proposed Newton-MUSIC effectively mitigates
the computational bottleneck inherent in exhaustive search over dense
grids by employing a coarse initialization followed by efficient iterative
refinement.

\subsection{Pilot Overhead Analysis \label{subsec:Pilot-Overhead-Analysis}}

A critical advantage of the proposed sensing-assisted channel estimation
framework is the drastic reduction in pilot overhead compared to conventional
pilot-based channel estimation methods. 

For conventional pilot-based channel estimation methods, acquiring
the full CSI matrix $\mathbf{H}\in\mathbb{C}^{N\times K}$ typically
necessitates sounding all $N$ potential antenna ports. Due to the
limited number of RF chains $M$, the BS must reconfigure the antenna
subset $\Omega$ across $\lceil N/M\rceil$ different configurations
to estimate the full channel. Assuming a total pilot budget of $T_{\textrm{p}}^{\textrm{(conv)}}$
is available, where $\text{(conv)}$ denotes conventional, the number
of pilot symbols allocated to each subset $\Omega$ is restricted
to $T_{\textrm{p}}^{\textrm{(conv)}}/\lceil N/M\rceil$. Denote $\mathbf{\Phi}_{\textrm{conv}}\in\mathbb{C}^{K\times\left(T_{\textrm{p}}^{\textrm{(conv)}}/\lceil N/M\rceil\right)}$
as the pilot matrix utilized for the conventional LS method. The LS
estimation problem for the activated subset channel $\hat{\mathbf{H}}_{\Omega}$
\cite{van1995channel}, given the received pilot signal $\mathbf{Y}_{\textrm{p}}^{\textrm{(conv)}}\in\mathbb{C}^{M\times\left(T_{\textrm{p}}^{\textrm{(conv)}}/\lceil N/M\rceil\right)}$,
is formulated as

\begin{equation}
\hat{\mathbf{H}}_{\Omega}=\arg\min_{\mathbf{H}_{\Omega}}\|\mathbf{Y}_{\textrm{p}}^{\textrm{(conv)}}-\mathbf{H}_{\Omega}\mathbf{\Phi}_{\textrm{conv}}\|_{F}^{2}.\label{eq:LSchannel}
\end{equation}
Consequently, the resulting NMSE $e_{\text{conv}}$ of the conventional
pilot-based channel estimation is given by
\begin{equation}
e_{\text{conv}}=\frac{M\sigma^{2}}{MT_{\textrm{p}}^{\text{(conv)}}/\lceil N/M\rceil}\approx\frac{N\sigma^{2}}{MT_{\textrm{p}}^{\text{(conv)}}}.\label{eq:econ}
\end{equation}

In stark contrast, the proposed framework strategically decomposes
the channel estimation task into three cascaded stages, as detailed
in Section \ref{subsec:Problem-Formulation}. Assuming $T_{\textrm{p}}^{\textrm{(prop)}}$
pilots are available, where $\text{(prop)}$ denotes proposed, the
resulting NMSE $e_{\text{prop}}$ of the proposed channel estimation
framework is approximately given by\footnote{It is important to acknowledge that the NMSE expression in \eqref{eq:epro}
represents a theoretical lower bound derived under the ideal assumption
of perfect DOA sensing (i.e., $\hat{\boldsymbol{\theta}}=\boldsymbol{\theta}$).
In practice, residual errors from the blind DOA sensing stage may
introduce slight performance degradation due to array manifold mismatch.} 
\begin{equation}
e_{\text{prop}}\approx\frac{K\sigma^{2}}{MT_{\textrm{p}}^{\text{(prop)}}}.\label{eq:epro}
\end{equation}

To evaluate the performance gain under an identical pilot overhead
constraint (i.e., setting $T_{p}^{(conv)}=T_{p}^{(prop)}$), the theoretical
NMSE reduction ratio $\eta$ achieved by the proposed framework is
derived as
\begin{equation}
\eta=\frac{e_{conv}}{e_{prop}}\approx\frac{N}{K}.\label{eq:ratio}
\end{equation}
This ratio demonstrates that in flexible-antenna architectures where
$N\gg K$, the proposed approach provides a substantial $\mathcal{O}(N/K)$
reduction in estimation error for a given pilot budget.

\section{Simulation Results \label{sec:Simulation-Results}}

In this section, we conduct simulations to compare the proposed Newton-MUSIC
algorithm with the existing benchmarks. We will first introduce simulation
settings and then discuss the results.

\subsection{System Settings}

We consider a multi-user flexible-antenna system with $N=40$ potential
antenna ports with a uniform half-wavelength spacing and $M=4$ RF
chains. The activated antenna subset follows a ULA configuration or
with $\Omega=\{1,2,3,4\}$ an MRA configuration \cite{moffet1968minimum}
with $\Omega=\{1,2,5,7\}$. We consider the general case of random
received signal power across users. For each Monte Carlo iteration,
which is set to $2000$, the path gains are generated as independent
and identically distributed (i.i.d.) samples $p_{k}\thicksim\mathcal{U}\left(p_{\min},p_{\max}\right)$
with $\frac{\max_{k}p_{k}}{\min_{k}p_{k}}\leq10$. The average received
signal-to-noise ratio (SNR) is defined as
\begin{equation}
\mathrm{SNR}\triangleq10\log_{10}\left(\frac{\frac{1}{K}\sum_{k=1}^{K}p_{k}}{\sigma^{2}}\right)\mathrm{dB}.\label{eq:SNR}
\end{equation}
For the non-Gaussian data symbols, we primarily consider QPSK modulation,
where symbols are drawn randomly from the alphabet set $S_{\textrm{QPSK}}=\left\{ a+jb\mid a,b\in\left\{ -\sqrt{2}/2,+\sqrt{2}/2\right\} \right\} $.
The source signals are assumed to be mutually independent and uncorrelated
with the additive noise, unless otherwise stated. The number of snapshots
$T$ is set as $2000$. For the classical MUSIC algorithm, the dense
grids are configured as $\Theta=\left\{ \theta_{\min}:\Delta_{\Theta}:\theta_{\max}\right\} $
with grid spacing $\Delta_{\Theta}=0.1\lyxmathsym{\textdegree}$.
For the proposed Newton MUSIC algorithm, the coarse grids are $\Theta_{\textrm{Newton}}=\left\{ \theta_{\min}:\Delta_{\Theta_{\textrm{Newton}}}:\theta_{\max}\right\} $
with $\Delta_{\Theta_{\textrm{Newton}}}=0.5\lyxmathsym{\textdegree}$.
The maximum number of Newton iterations is set to $Z=200$. 

For the performance comparison of DOA sensing, we consider the following
algorithms: 
\begin{itemize}
\item \textbf{SOC-MUSIC}: The FOC-based MUSIC algorithm following the settings
from \cite{schmidt1986multiple};
\item \textbf{SOC-Newton-MUSIC}: The difference co-array SOC-based Newton-MUSIC
algorithm proposed in this paper; 
\item \textbf{FOC-MUSIC}: The difference co-array FOC-based classical MUSIC
algorithm; 
\item \textbf{FOC-Newton-MUSIC}: The difference co-array FOC-based Newton-MUSIC
algorithm proposed in this paper; 
\item \textbf{FOC-MUSIC}-\textbf{Sum}: The sum co-array FOC-based classical
MUSIC algorithm from \cite{peng2024under}.
\begin{figure}[t]
\begin{centering}
\subfloat[ULA with SOC-based methods.]{\includegraphics[scale=0.24]{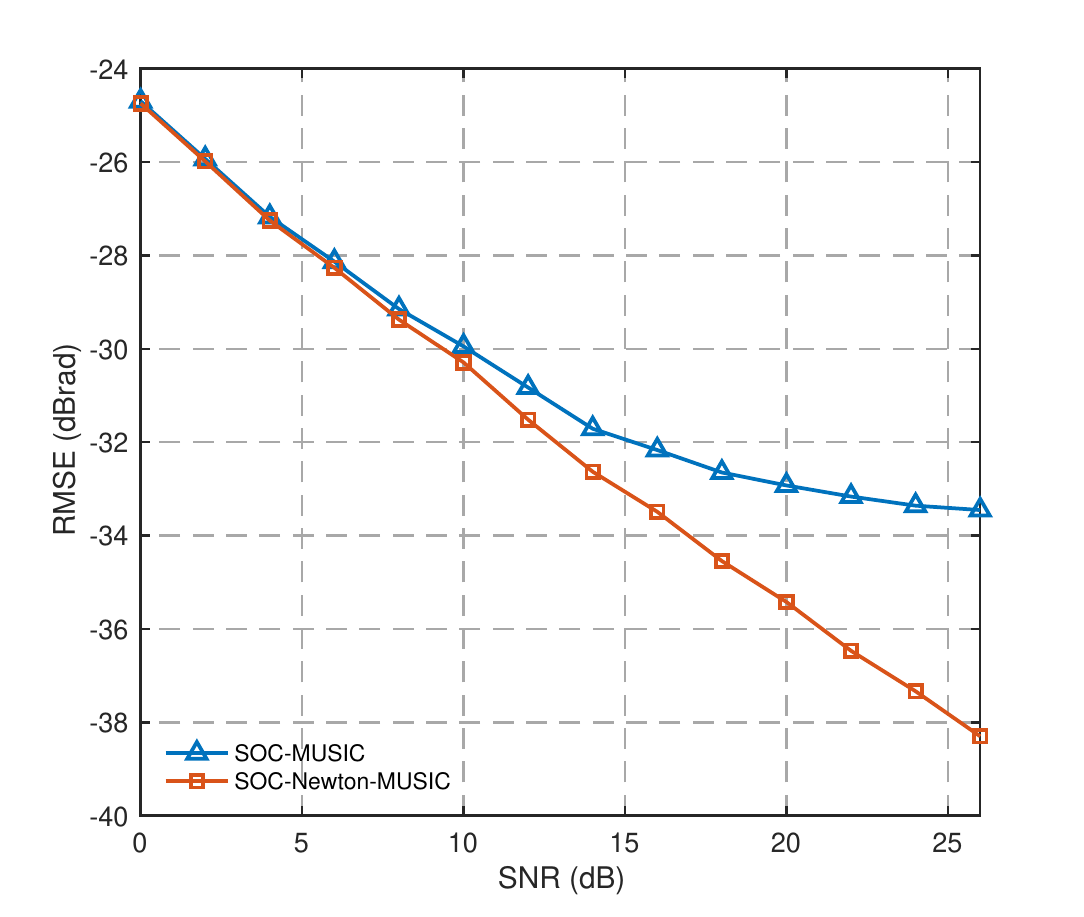}

}\subfloat[MRA with FOC-based methods.]{\includegraphics[scale=0.24]{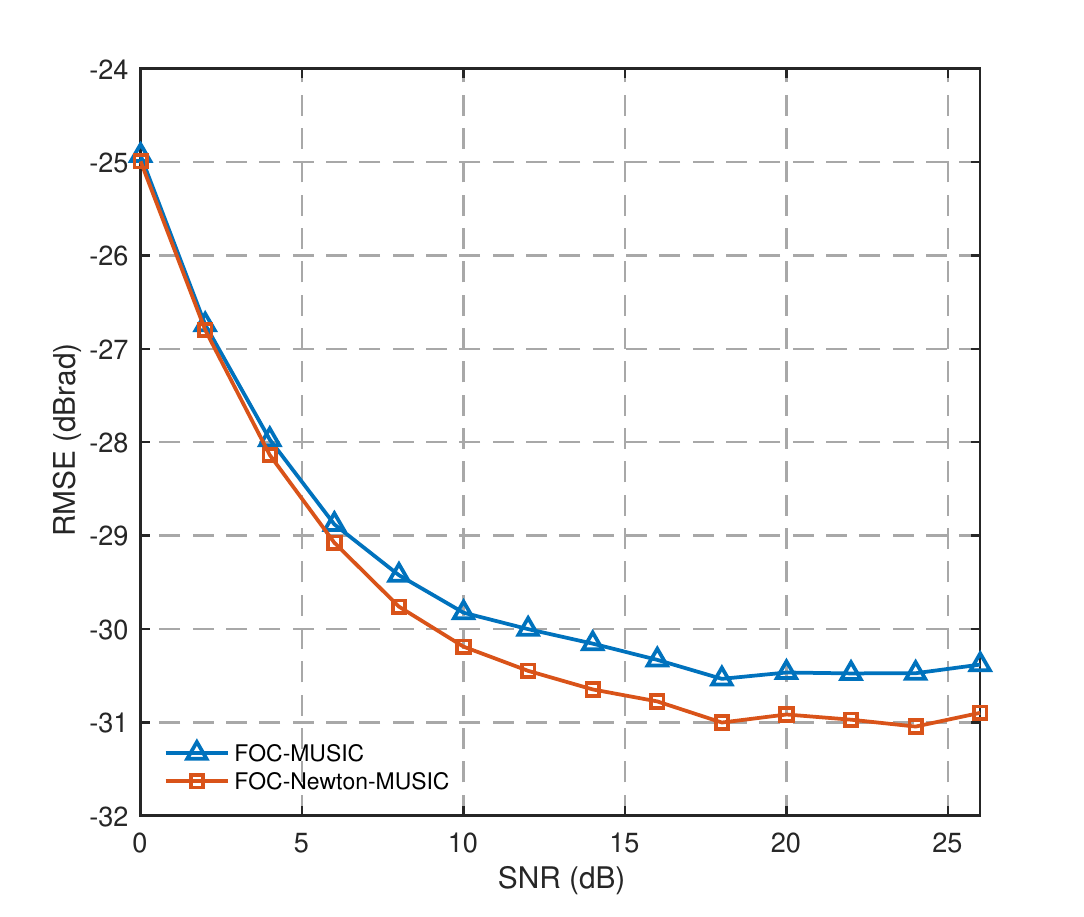}

}
\par\end{centering}
\caption{RMSE of the DOA sensing for different configurations.\label{fig:DOA_Sensing}}
\end{figure}
\begin{figure}[t]
\begin{centering}
\subfloat[ULA with SOC-based methods.]{\includegraphics[scale=0.24]{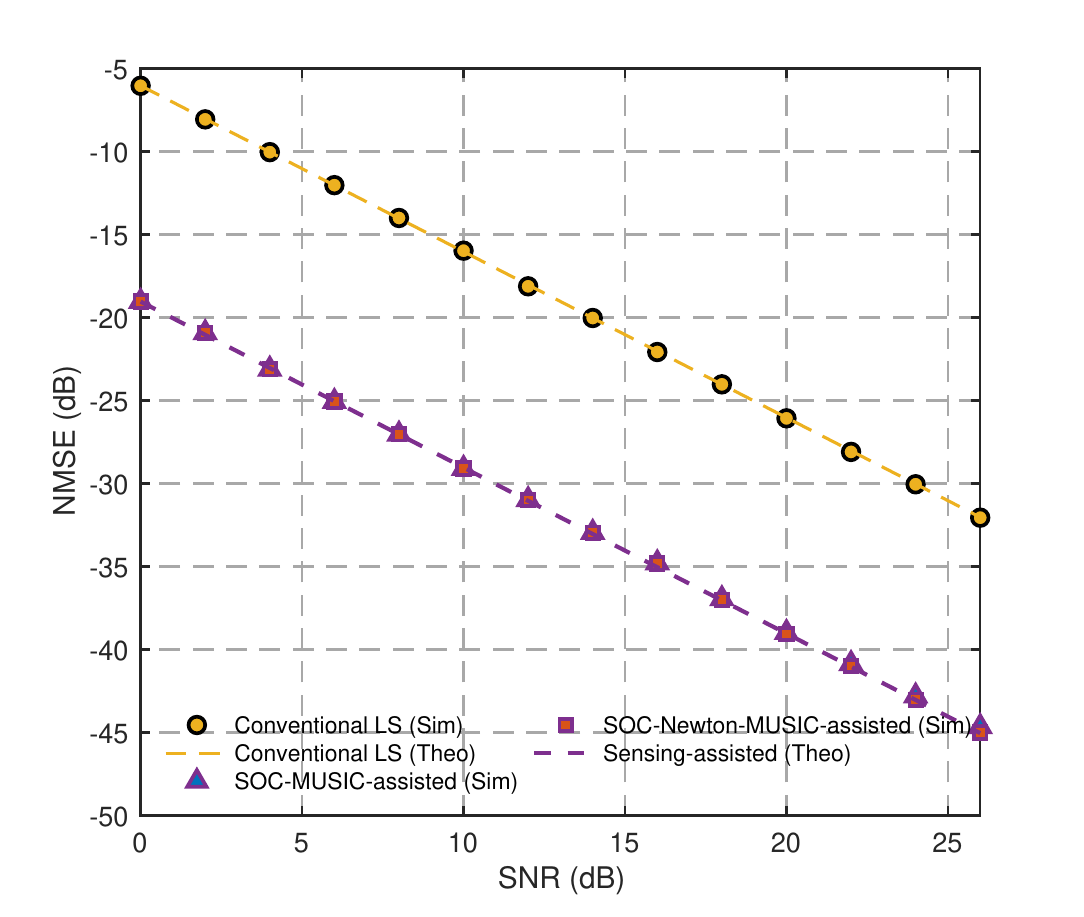}

}\subfloat[MRA with FOC-based methods.]{\includegraphics[scale=0.24]{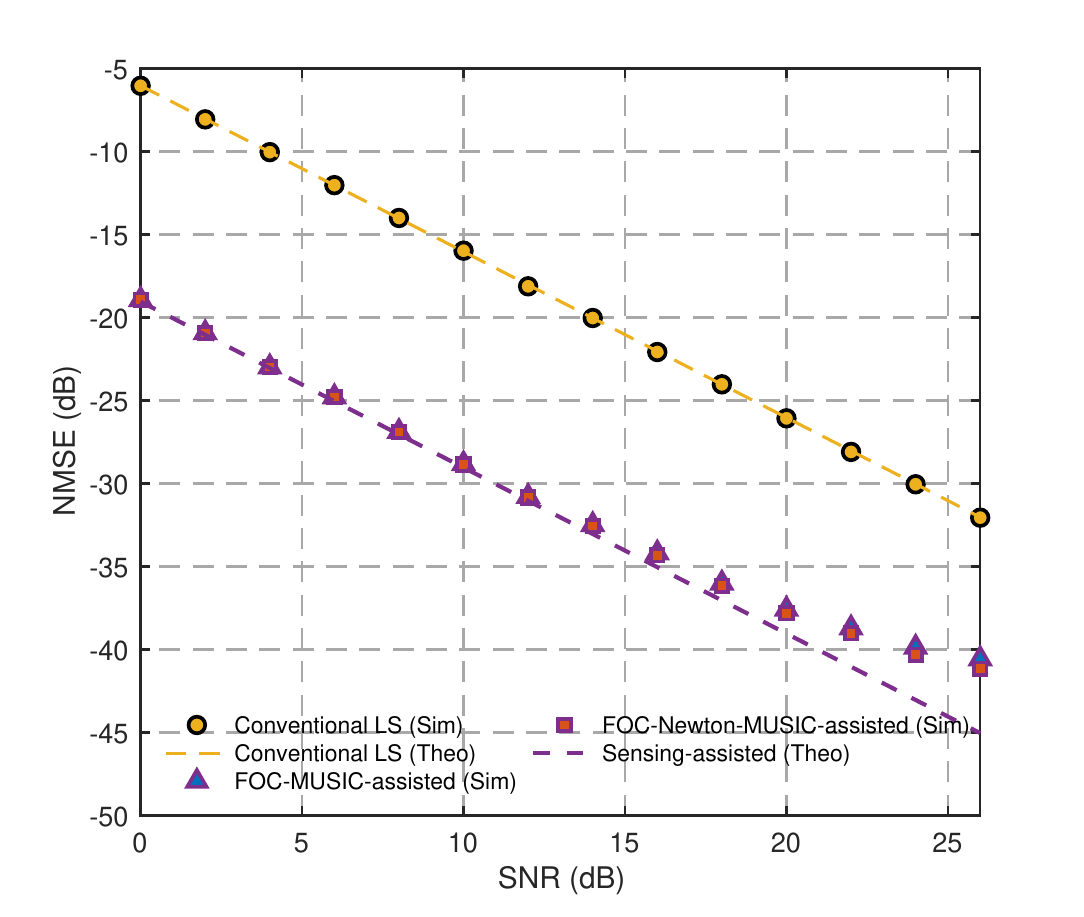}

}
\par\end{centering}
\caption{Channel estimation performance for for different configurations. \label{fig:Channel_Estimation}}
\end{figure}
\end{itemize}

\subsection{Channel Estimation }

To ensure a fair comparison under a fixed overhead, the total pilot
budget is set to $T_{\textrm{p}}=40$. Following the analysis in Section
\ref{subsec:Pilot-Overhead-Analysis}, the proposed framework dedicates
all available pilots to the single activated subset. In contrast,
the conventional pilot-based method must distribute this total budget
across $\lceil N/M\rceil=10$ sequential configurations, restricting
the available pilots per activated subset to $T_{\textrm{p}}^{\textrm{(conv)}}/\lceil N/M\rceil=4$.
To ensure a fair comparison across different SNR levels, for a given
number of sources $K=2$, the DOAs $\boldsymbol{\theta}$ are set
as $\left\{ -20\lyxmathsym{\textdegree}+\Delta_{1}^{\textrm{(per)}},-25\lyxmathsym{\textdegree}+\Delta_{2}^{\textrm{(per)}}\right\} $,
where $\Delta_{k}^{\textrm{(per)}}\thicksim\mathcal{U}\left(-0.5\lyxmathsym{\textdegree},0.5\lyxmathsym{\textdegree}\right)$
is an introduced small random perturbation.

Fig. \ref{fig:DOA_Sensing} illustrates the RMSE of the DOA sensing
performance versus SNR for the SOC-based ULA configuration and the
FOC-based MRA configuration, respectively. In the low-SNR regime,
the proposed Newton-MUSIC variants achieve estimation accuracy comparable
to conventional MUSIC algorithms. However, as the SNR increases, the
classical grid-based MUSIC algorithms exhibit a noticeable performance
plateau. This bottleneck is primarily caused by the inherent grid
quantization error, as the true DOAs generally do not fall exactly
on the predefined discrete angular grids. In stark contrast, the proposed
SOC-Newton-MUSIC effectively circumvents this limitation. By applying
iterative Newton refinements, our method significantly suppresses
the off-grid mismatches, thereby continuously reducing the RMSE at
high SNRs and demonstrating superior high-resolution capability. Similarly,
for the FOC-based MRA configuration, the proposed FOC-Newton-MUSIC
algorithm yields a consistent performance improvement by mitigating
the grid-induced quantization bounds. More importantly, these accuracy
enhancements are achieved alongside significant computational efficiency.
The proposed Newton-MUSIC framework reduces the computational overhead
in the blind DOA sensing stage by approximately two-thirds compared
to classical MUISC algorithms.

The advantages acquired in the sensing stage naturally extend to the
communication stage, as verified by the NMSE of the channel estimation
in Fig. \eqref{fig:Channel_Estimation}. It can be observed that all
sensing-assisted channel estimation frameworks significantly outperform
the conventional LS method across the entire SNR range. This substantial
performance margin stems from the fact that our proposed framework
explicitly exploits the sparse geometric structure of the channel
(i.e., DOAs sensing) rather than blindly estimating an unstructured,
high-dimensional channel matrix. Moreover, benefiting from the highly
accurate DOA inputs provided by the Newton refinements, both the SOC-Newton-MUSIC-assisted
and FOC-Newton-MUSIC-assisted approaches achieve the lowest NMSE in
their respective array configurations. Finally, the Monte Carlo simulation
markers of the proposed SOC-Newton-MUSIC-assisted scheme exhibit a
tight alignment with the theoretical NMSE curves. While the other
methods closely track the theoretical bounds in the low-SNR regime,
they begin to deviate at higher SNRs due to the aforementioned grid
quantization errors. This close agreement validates the robustness
and reliability of the preceding DOA sensing stage.%
\begin{figure}[t]
\begin{centering}
\subfloat[NLOS environments.]{\includegraphics[scale=0.24]{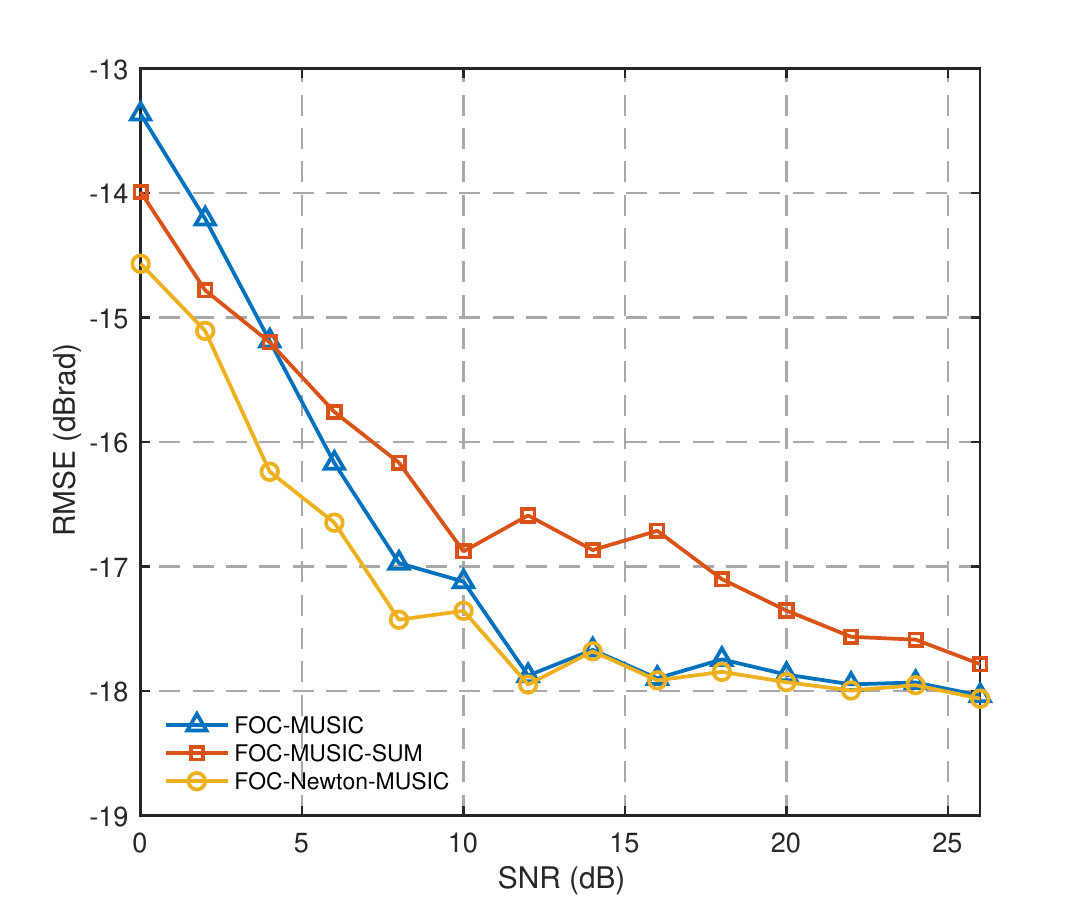}

}\subfloat[Underdetermined regimes.]{\includegraphics[scale=0.24]{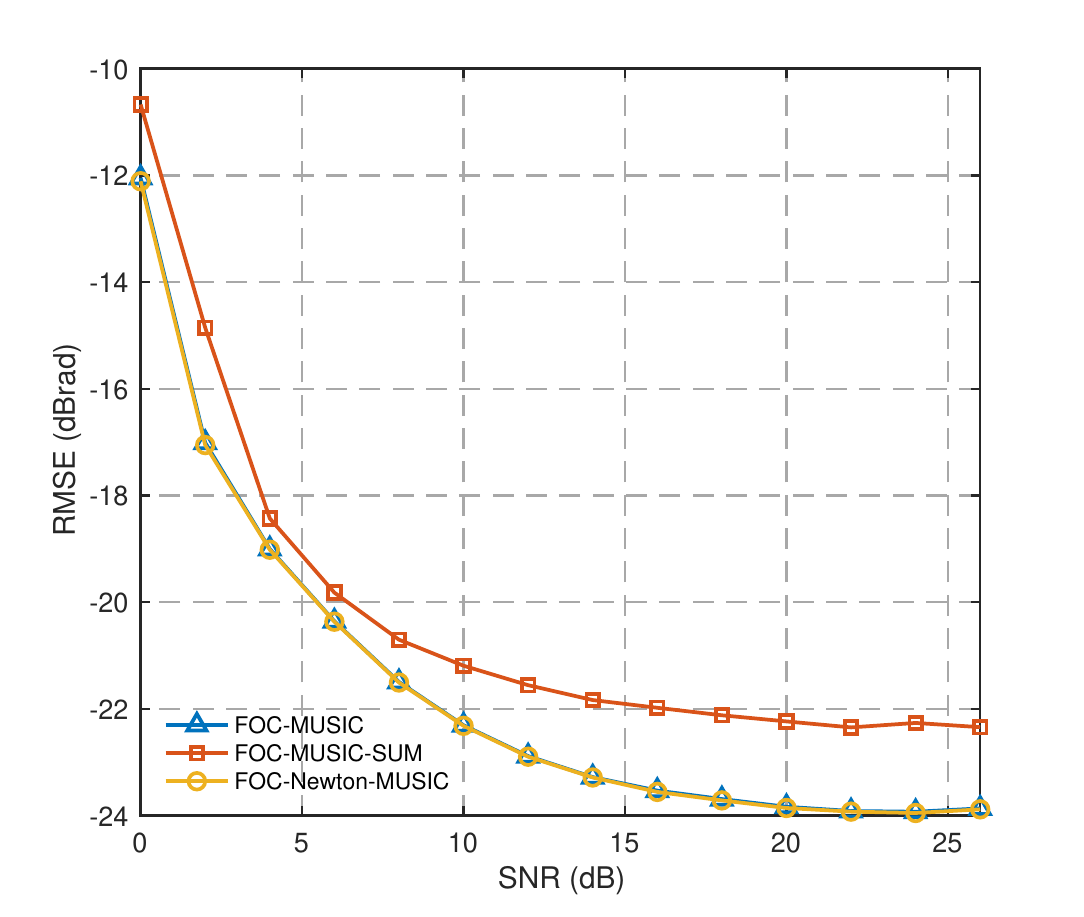}

}
\par\end{centering}
\caption{RMSE of DOA sensing versus SNR under the MRA configuration. \label{fig:DOA_senario}}
\end{figure}

\subsection{NLOS Environments with Coherent Multipath }

This simulation considers $K=3$ sources with $s_{2}=0.9s_{1}$ (coherent)
and $s_{3}$ (independent). The DOAs $\boldsymbol{\theta}$ are located
at $\left\{ -45\lyxmathsym{\textdegree}+\Delta_{1}^{\textrm{(per)}},0\lyxmathsym{\textdegree}+\Delta_{2}^{\textrm{(per)}},40\lyxmathsym{\textdegree}+\Delta_{3}^{\textrm{(per)}}\right\} $.
The MRA configuration is employed.

Fig. \ref{fig:DOA_senario}(a) illustrates the RMSE of DOA sensing
versus SNR for the NLOS environments with coherent multipath. As depicted,
the FOC-based algorithms maintain robust sensing performance even
in the presence of highly coherent multipath, a challenging environment
where conventional SOC-based methods typically fail due to the rank-deficiency
of the SOC. Among the evaluated algorithms, the proposed FOC-Newton-MUSIC
consistently achieves the superior accuracy across the entire SNR
range. By replacing the exhaustive search over dense grids with iterative
Newton-based refinements, our method effectively circumvents the grid
quantization error floor that limits the resolution of standard FOC-MUSIC
and FOC-MUSIC-SUM. Specifically, the FOC-Newton-MUSIC demonstrates
a significant precision gain. This consistent performance profile
substantiates the robustness and high-resolution capability of the
proposed Newton-MUSIC algorithm in addressing the inherent limitations
of subspace-based sensing for coherent signals. Simultaneously, since
the parameter settings governing complexity remain unchanged in this
scenario, our proposed algorithm continues to maintain the lowest
complexity. Furthermore, it is observed that the FOC-MUSIC-SUM algorithm
exhibits degraded performance compared to the standard difference
co-array-based FOC-MUSIC in this NLOS environments. This instability
arises because the sum co-array inherently contains holes, which degrades
the spatial resolution capabilities and restricts the overall algorithmic
robustness when resolving highly correlated signals.

It is also worth noting that the simulated RMSE curves exhibit slight
fluctuations, even when averaged over thousands of Monte Carlo runs.
This phenomenon stems from an inherent limitation in the RMSE statistical
evaluation for MUSIC algorithms. Specifically, in challenging NLOS
environments, these algorithms occasionally suffer from missed detections,
where the number of identified spatial peaks is less than the actual
number of sources $K$. To ensure a rigorous evaluation, such missed
detection events are penalized by assigning a maximum angular error
of $180\lyxmathsym{\textdegree}$ to the unresolved sources. Consequently,
these extreme outlier values induce a noticeable variance in the overall
RMSE calculation, resulting in the observed curve fluctuations that
cannot be completely eliminated by merely increasing the number of
Monte Carlo trials. %

\subsection{Underdetermined Regime}

The number of sources is set to $K=6$ to establish the underdetermined
regime. The DOAs $\boldsymbol{\theta}$ are configured as $\left\{ -55\lyxmathsym{\textdegree}+\Delta_{1}^{\textrm{(per)}},\right.$$-32\lyxmathsym{\textdegree}+\Delta_{2}^{\textrm{(per)}},-10\lyxmathsym{\textdegree}+\Delta_{3}^{\textrm{(per)}},10\lyxmathsym{\textdegree}+\Delta_{4}^{\textrm{(per)}},32\lyxmathsym{\textdegree}+\Delta_{5}^{\textrm{(per)}},$$\left.55\lyxmathsym{\textdegree}+\Delta_{6}^{\textrm{(per)}}\right\} $.
The MRA configuration is adopted in this simulation.

Fig. \ref{fig:DOA_senario}(b) illustrates the RMSE of DOA sensing
versus SNR in the underdetermined regime. It can be observed that
all evaluated FOC-based frameworks are capable of resolving the distinct
DOAs, corroborating the theoretical premise that higher-order statistics
synthetically expand the virtual array aperture and enhance the available
DOF far beyond the physical hardware constraints. Among these capable
methods, the proposed FOC-Newton-MUSIC algorithm achieves estimation
accuracy comparable with the standard FOC-MUSIC baseline across the
entire SNR spectrum. Although the Newton refinement does not yield
a distinct accuracy advantage in this specific underdetermined simulation,
it is worth noting that our proposed approach attains this robust
performance while obviating the need for an exhaustive search over
dense grids. This preserves its key advantage of maintaining a substantially
lower computational burden relative to the other evaluated techniques.
Furthermore, it is observed that the FOC-MUSIC-SUM approach exhibits
suboptimal performance compared to the standard difference co-array-based
FOC-MUSIC. This performance degradation arises primarily because the
sum co-array generates a smaller virtual array size compared to the
difference co-array, thereby providing fewer spatial DOFs, which are
critical for robust estimation in severely underdetermined regimes.
Consequently, this performance profile indicates that the proposed
Newton-enhanced framework effectively mitigates the physical DOF limitations
of underdetermined systems, achieving reliable spatial resolution
and robustness alongside significantly improved computational efficiency.
\begin{figure}[t]
\begin{centering}
\subfloat[$\Delta_{\min}=20\lyxmathsym{\protect\textdegree}$.]{\includegraphics[scale=0.24]{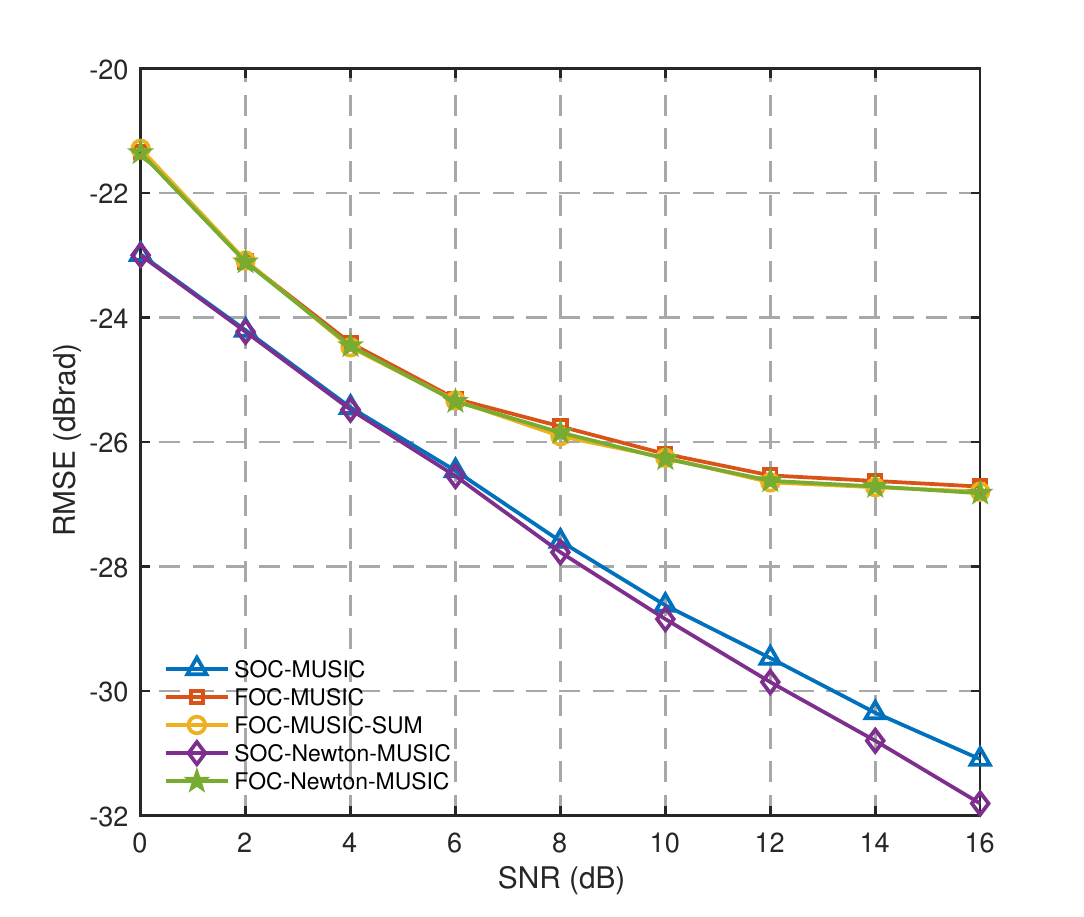}

}\subfloat[$\Delta_{\min}=3\lyxmathsym{\protect\textdegree}$.]{\includegraphics[scale=0.24]{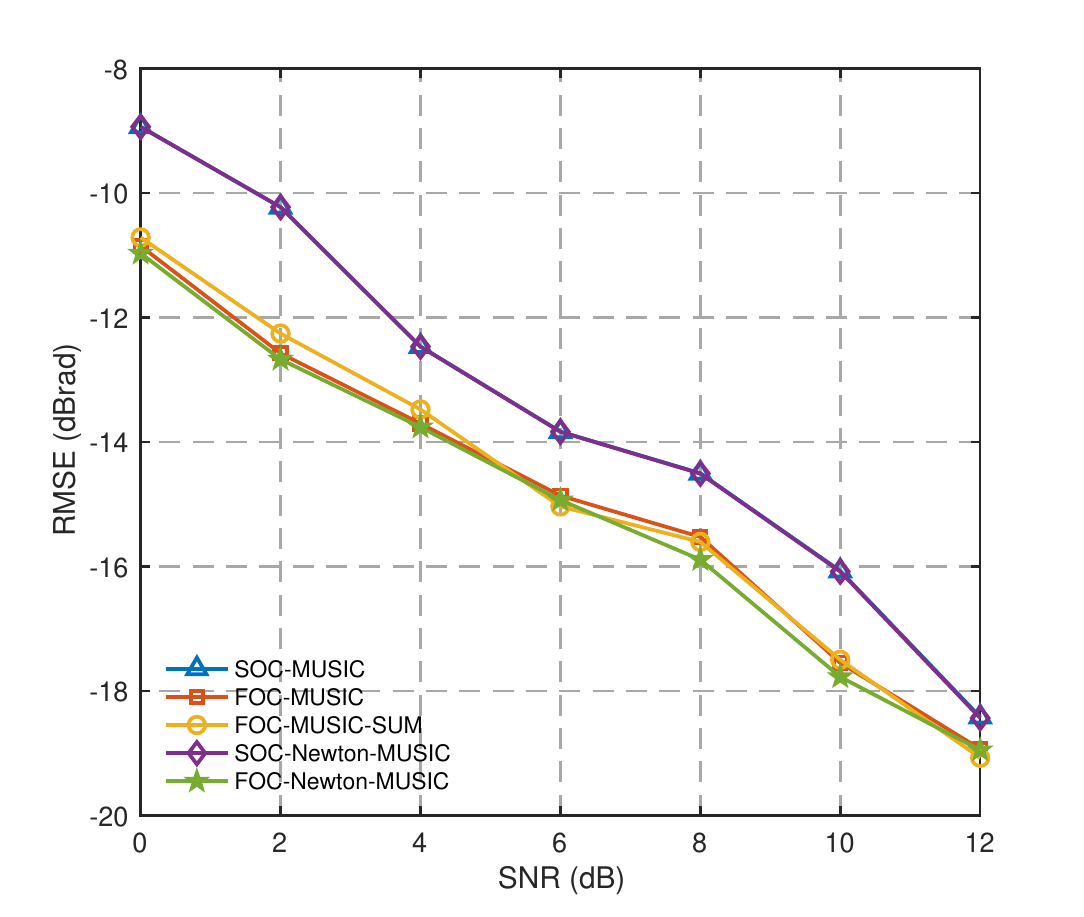}

}
\par\end{centering}
\caption{RMSE of the DOA sensing versus SNR for random sources with different
minimum angular separations. \label{fig:DOA_spacing}}
\end{figure}

\subsection{Random DOAs with Different Separation}

To validate the general applicability of the proposed unified framework
under arbitrary spatial conditions, we evaluate its performance using
randomly generated DOAs, including an extreme scenario with tightly
spaced sources. For a given number of sources $K=2$, the DOA $\left\{ \theta_{k}\right\} _{k=1}^{K}$
are generated uniformly over the angular range $\left[\theta_{\min},\theta_{\max}\right]$,
subject to a minimum angle separation constraint defined as $\Delta_{\min}\triangleq\min_{k_{1}\neq k_{2}}\left|\theta_{k_{1}}-\theta_{k_{2}}\right|,\forall k_{1},k_{2}\in\{1,2,\ldots,K\}$.
In this simulations, we set $\theta_{\min}=-60\lyxmathsym{\textdegree}$,
$\theta_{\max}=60\lyxmathsym{\textdegree}$, utilizing the ULA configuration.

Fig. \ref{fig:DOA_spacing} illustrates the RMSE of DOA estimation
versus SNR for random sources with different minimum angular separations.
Fig. \ref{fig:DOA_spacing}(a) depicts the performance with a sufficiently
large separation ($\Delta_{\min}=20\lyxmathsym{\textdegree}$). The
results confirm that the proposed unified framework is highly effective
and generalizable for randomly distributed DOAs, maintaining stable
estimation accuracy. Consistent with previous observations, the SOC-Newton-MUSIC
effectively mitigate the grid quantization floor at higher SNRs.

Fig. \ref{fig:DOA_spacing}(b) further investigates the extreme scenario
where the sources are located in small separation ($\Delta_{\min}=3\lyxmathsym{\textdegree}$).
As the spatial interval shrinks, the conventional SOC-based methods
experience severe performance degradation. This is primarily because
their wider mainlobes struggle to distinguish highly proximate spatial
signatures. In contrast, all FOC-based algorithms demonstrate significantly
enhanced robustness, maintaining much lower RMSE levels. This advantage
stems from the synthetic aperture expansion of higher-order statistics,
which yields narrower mainlobes and intrinsically higher spatial resolution.
Consequently, the proposed unified framework ensures reliable spatial
resolution for challenging closely spaced targets, but with a substantially
reduced computational burden, underscoring its practical versatility
and efficiency.

\section{Conclusion \label{sec:Conclusion}}

In this paper, we developed a unified sensing-assisted channel estimation
framework for flexible-antenna systems, driven by three key insights.
First, we showed that decoupling DOA sensing from dedicated pilots
and extracting spatial information directly from the data payload
drastically reduces the prohibitive pilot overhead. Second, the spatial
DOFs can be structurally expanded without additional hardware. By
leveraging the FOC-based difference co-array, high-resolution source
identifiability is inherently restored even in challenging underdetermined
and coherent multipath regimes. Third, our analysis demonstrated that
reformulating the spatial spectrum search as a continuous optimization
problem is essential. Continuous Newton refinements successfully bypass
the fundamental resolution limits and grid quantization errors that
typically cause performance plateaus in classical subspace methods.
While this paper has established a unified framework, future work
will focus on evaluating its robustness against realistic non-idealities.
Specifically, extending the proposed algorithms to account for hardware
impairments, near-field propagation, and Doppler effects represents
a critical next step toward practical deployment.

\bibliographystyle{IEEEtran}
\bibliography{reference}

\end{document}